\documentclass[aps,prl,preprint,groupedaddress]{revtex4-1}

\usepackage{standalone}
\usepackage{graphicx}
\usepackage{subfigure}
\usepackage{array}
\usepackage{xr}
\newcommand{\PreserveBackslash}[1]{\let\temp=\\#1\let\\=\temp}
\newcolumntype{C}[1]{>{\PreserveBackslash\centering}p{#1}}
\newcolumntype{R}[1]{>{\PreserveBackslash\raggedleft}p{#1}}
\newcolumntype{L}[1]{>{\PreserveBackslash\raggedright}p{#1}}

\begin{document}

\title{First-principles prediction of a new family of layered topological insulators}

\author{Ping Li}
\email[]{pingli.phys@hotmail.com}
\affiliation{School of Mathematics and Physics, Anhui Jianzhu University, Hefei, 230601, China}

\author{Jiangying Yu}
\affiliation{School of Mathematics and Physics, Anhui Jianzhu University, Hefei, 230601, China}
\author{Jinrong Xu}
\affiliation{School of Mathematics and Physics, Anhui Jianzhu University, Hefei, 230601, China}
\author{Li Zhang}
\affiliation{School of Mathematics and Physics, Anhui Jianzhu University, Hefei, 230601, China}
\author{Kai Huang}
\affiliation{School of Mathematics and Physics, Anhui Jianzhu University, Hefei, 230601, China}
\date{\today}

\begin{abstract}
Dozens of layered V$_2$IV$_2$VI$_6$ (V=P, As, Sb, Bi, IV=Si, Ge, Sn, Pb, VI=S, Se, Te) materials are investigated, several of which have been successfully synthesized in experiment. Among them, we predict nine strong topological insulators (TIs), two strong topological metals (TMs) and nearly twenty trivial insulators at their equilibrium structures. The TIs are in the (1;111) topological class, with energy gaps ranging from 0.04 to 0.2 eV. The strong TMs and the trivial insulators belong to the (1;111) and (0;000) topological classes, respectively. Small compressive strains easily turn some of the trivial insulators into strong TIs. This study enriches not only the family of topological materials but also the family of van der Waals layered materials, providing promising candidates for the future spintronic devices.
\end{abstract}
\pacs{}
\maketitle

\section{Introduction}

About a decade ago, topological insulators (TIs) were theoretically predicted \cite{prl106803, prb121306, prb195322, prb045302} and experimentally realized in real materials \cite{np398, np438}. Since then, TIs have attracted tremendous attention due to their fundamental interest as well as potential applications \cite{rmp3045,rmp1057}. A great many three-dimensional (3D) TIs \cite{prl236402, jpcl332, jmcc4752, prb085116, sc2199, prl036404, prl146801, prl206803} have been proposed, but only a few of them have been experimentally synthesized and veryfied \cite{prl036404, prl146801, prl206803}, among which only the TIs of the Bi$_2$Se$_3$ family have been widely and intensively studied. This family of TIs have large energy gaps (typically, 0.3 eV for Bi$_2$Se$_3$), enough for room temperature utilizations; they are van der Waals (vdW)-layered, so defective surface states can be largely avoided and the lattice matching requirement can be significantly reduced in epitaxial growth on substrates. However, these TIs suffer from intrinsic doping. Bi$_2$Se$_3$ and Bi$_2$Te$_3$ are usually $n$-type \cite{prl146401, science178, np398} while Sb$_2$Te$_3$ shows $p$-type \cite{prl146401}. Though the Fermi levels can be tuned by extrinsic doping or alloying \cite{prb165311, prb155301, nc574}, such experiments must be carefully controlled and the studies are still on the way. Searching or predicting new TIs, especially those with large energy gaps and layered crystal structures, are sitll of great interest.

Recently, ternary compounds Bi$_2$Si$_2$Te$_6$ and Sb$_2$Si$_2$Te$_6$ have been successfully synthesized in experiment \cite{dis2010}. These compounds are vdW-layered, belonging to the V$_2$IV$_2$VI$_6$ (V=P, As, Sb, Bi, IV=Si, Ge, Sn, Pb, VI=S, Se, Te) family. Here we show that Bi$_2$Si$_2$Te$_6$ and Sb$_2$Si$_2$Te$_6$ can be easily turned into strong TIs by small compressive strains. Moreover, tens of the V$_2$IV$_2$VI$_6$ family of compounds are intensively investigated. Based on the calculated $Z_2$ invariants ($\nu_0; \nu_1\nu_2\nu_3$), we predict nine strong TIs, two strong topological metals (TMs), and nearly twenty trivial insulators  at equilibrium lattice constants. More strong TIs and TMs can be obtained by applying small compressive strains to the trivial insulators. The TIs have energy gaps ranging from several tens of meV to 0.2 eV, and support one single Dirac cone at the topological surface states.

\section{Crystal Structure and Method}
\begin{figure*}
\includegraphics [width=12cm] {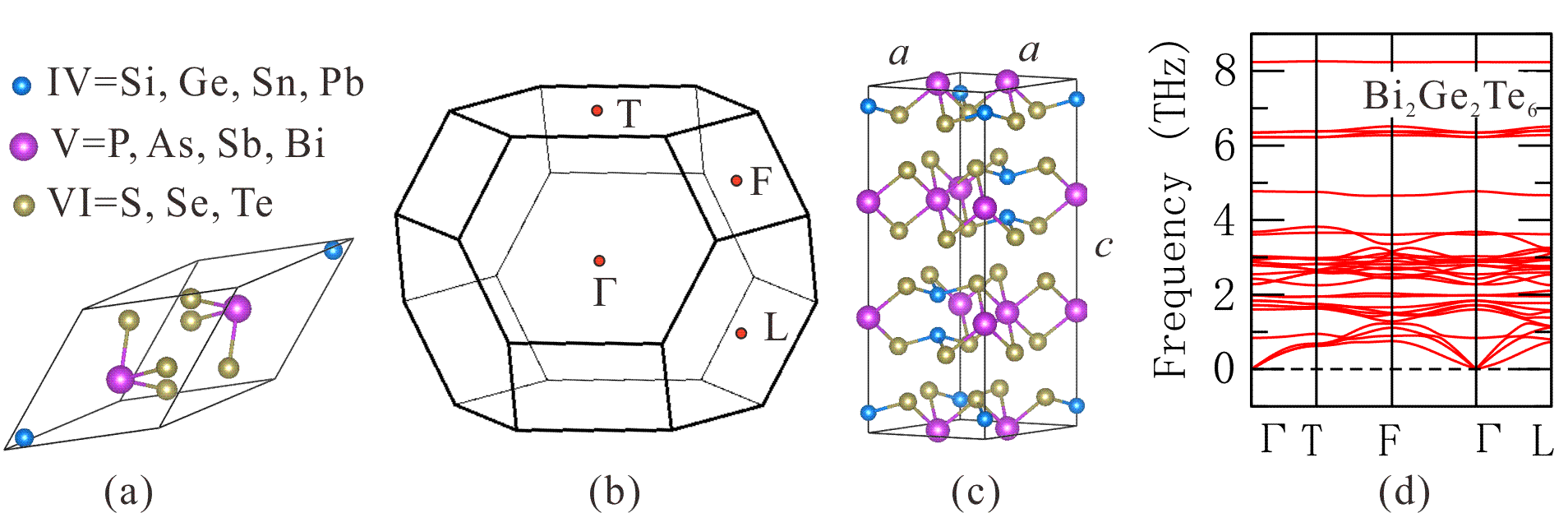}
\caption{(a) Primitive cell, (b) Brillouin zone and (c) conventional cell of the V$_2$IV$_2$VI$_6$ family of crystals. (d) Phonon spectrum of Bi$_2$Ge$_2$Te$_6$.}
\label{crystal}
\end{figure*}

The V$_2$IV$_2$VI$_6$ (V=P, As, Sb, Bi, IV=Si, Ge, Sn, Pb, VI=S, Se, Te) family of compounds share the same rhombohedral structure with a space group of $R\overline{3}$ (No. 148) with ten atoms in each primitive cell. The atomic structure and the Brillouin zone of the primitive cell are shown in Fig.~\ref{crystal} (a) and (b), respectively. Along the trigonal axis or the [111] direction, VI$_3$-IV-V$_2$-IV-VI$_3$ ten-atom layers are periodically stacked at five different heights, forming quintuple layers (QLs) as Fig.~\ref{crystal} (c) shows. The conventional cell contains 30 atoms and three QLs. The inter-QL coupling is strong, while the intra-QL coupling is much weaker, forming vdW-type interactions, similar to the Bi$_2$Se$_3$ case \cite{np438}.

This work is performed using the first-principles method within the framework of density functional theory of generalized
gradient approximations (DFT-GGA) \cite{prb864, prl3865} at \it{High} \rm precision, as implemented in the VASP codes \cite{prb11169}. $8\times8\times8$ and $8\times8\times2$ K-point meshes are used for primitive and conventional cells, respectively. The atoms and the cell size are fully relaxed until the force acting on each ion is less than 0.01 eV/\r{A}. Surface states are calculated within slab supercells containing nine QLs of V$_2$IV$_2$VI$_6$ and a vacuum thicker than 20 \r{A}. Phonon spectra are obtained by the calculation of the interatomic force constants (IFCs) in VASP, with the help of the Phonopy package \cite{sm1}. The $Z_2$ topological invariants ($\nu_0; \nu_1\nu_2\nu_3$) \cite{prl106803, prb121306, prb195322} are calculated using the parity method proposed by Fu and Kane \cite{prb045302}.

\section{Results}
\begin{figure*}[thp]
\includegraphics [width=9cm] {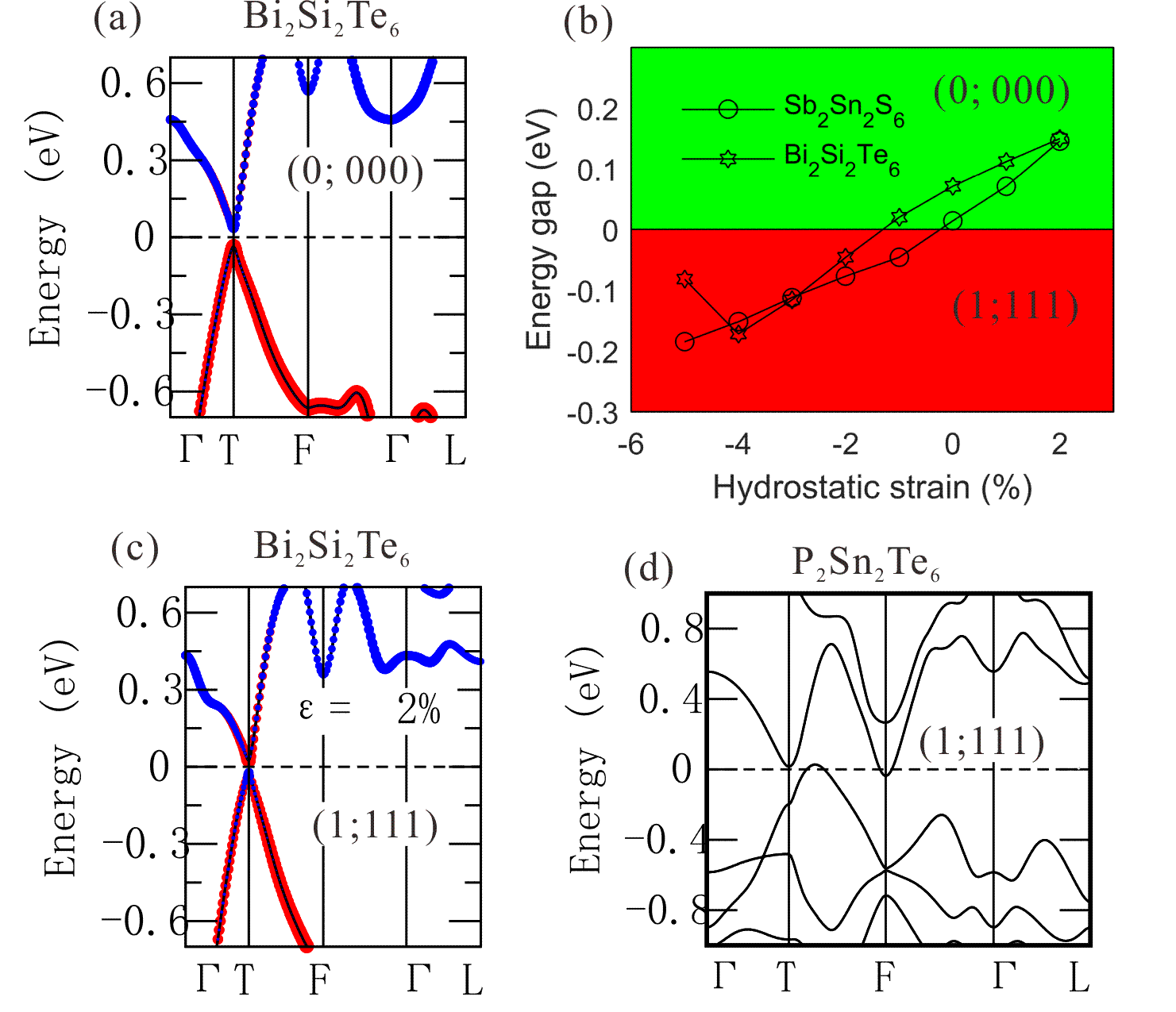}
\caption{(a) Band structure of Bi$_2$Si$_2$Te$_6$. (b) Energy gaps at different hydrostatic strains of Bi$_2$Si$_2$Se$_6$ and Sb$_2$Sn$_2$S$_6$, where green and red colors indicate topological trivial and nontrivial regions, respectively. (c) Band structure of Bi$_2$Si$_2$Te$_6$ at a 2\% compressive strain. (d) Band structure of P$_2$Sn$_2$Te$_6$. Red and blue symbols in (a) and (c) denote the Te-p and the Bi-p states, respectively, and the symbol size indicates the contribution weight. (0;000) and (1;111) are the $Z_2$ topological invariants.}
\label{band2}
\end{figure*}

We first consider Bi$_2$Si$_2$Te$_6$ and Sb$_2$Si$_2$Te$_6$, which have been successfully synthesized in experiment \cite{dis2010, dis2014}. The calculated lattice constants of Bi$_2$Si$_2$Te$_6$ (Sb$_2$Si$_2$Te$_6$) are $a=7.317$ \r{A}, $c=21.438$ \r{A} ($a=7.223$ \r{A}, $c=21.236$ \r{A}), in good agreement with the experimental data $a=7.269$ \r{A}, $c=21.293$ \r{A} ($a=7.169$ \r{A}, $c=21.186$ \r{A}) \cite{dis2014}. The band structure of Bi$_2$Si$_2$Te$_6$ is shown in Fig.~\ref{band2} (a) as an example. Its valence band maximum (VBM) and conduction band minimum (CBM) are located at T point in the Brillouin zone, forming a direct energy gap of 0.07 eV. The VBM and the CBM are mainly contributed by the Te-p and the Bi-p states, respectively. The band structure of Sb$_2$Si$_2$Te$_6$ is similar, but having a larger direct gap of 0.15 eV. According to the calculated $Z_2$ topological invariants, both materials are in the (0;000) topological class, so they are trivial insulators.

Strains can remarkably change the band structure of Bi$_2$Si$_2$Te$_6$ and Sb$_2$Si$_2$Te$_6$ as well as their topological nature. Fig.~\ref{band2} (b) shows the gap-dependence of Bi$_2$Si$_2$Te$_6$ on the hydrostatic strain $\varepsilon$, where negative and positive $\varepsilon$ values represent compressive and tensile strains, respectively. A small compressive strain of 2\% is enough to turn the material into a strong TI of the (1;111) topological class. The band structure of Bi$_2$Si$_2$Te$_6$ at $\varepsilon=-2\%$ is shown in Fig.~\ref{band2} (c). The CBM and the VBM are mainly composed of the Bi-p and Te-p states, respectively. A band inversion obviously occurs at T point, consistent with the topological phase transition.

Sb$_2$Si$_2$Te$_6$ behaves differently under strains. Compressive strains less than 2\% cannot change the topological nature of Sb$_2$Si$_2$Te$_6$. At $\varepsilon=-3\%$, the VBM remains at T point, while the CBM shifts from T to F point and becomes energetically lower than the VBM, making the material metallic. However, local energy gaps exist everywhere around the Fermi level in the Brillouin zone. The calculated $Z_2$ invariants indicate the material now belongs to the (1;111) topological class, i.e., it becomes a strong TM.

Studies are extended to dozens of similar compounds, including P$_2$Sn$_2$Se$_6$, P$_2$Sn$_2$Te$_6$, and 27 V$_2$IV$_2$VI$_6$ materials with V=As, Sb, Bi, IV=Si, Ge, Sn, VI=S, Se, Te. Phonon dispersions of all these crystals are calculated, and no imaginary modes are found, indicating they are all dynamically stable. However, when several compounds containing Pb atoms including Bi$_2$Pb$_2$Se$_6$ and Bi$_2$Pb$_2$Te$_6$ are considered, they show obviously imaginary modes, thus they are dynamically unstable. The phonon dispersions of Bi$_2$Ge$_2$Te$_6$ are shown in Fig.~\ref{crystal} (d) as an example, while those of more compounds can be found in Fig.~\ref{ext1-spho} in the Supplementary Materials.

\begin{figure*}[thp]
\includegraphics [width=8.5cm] {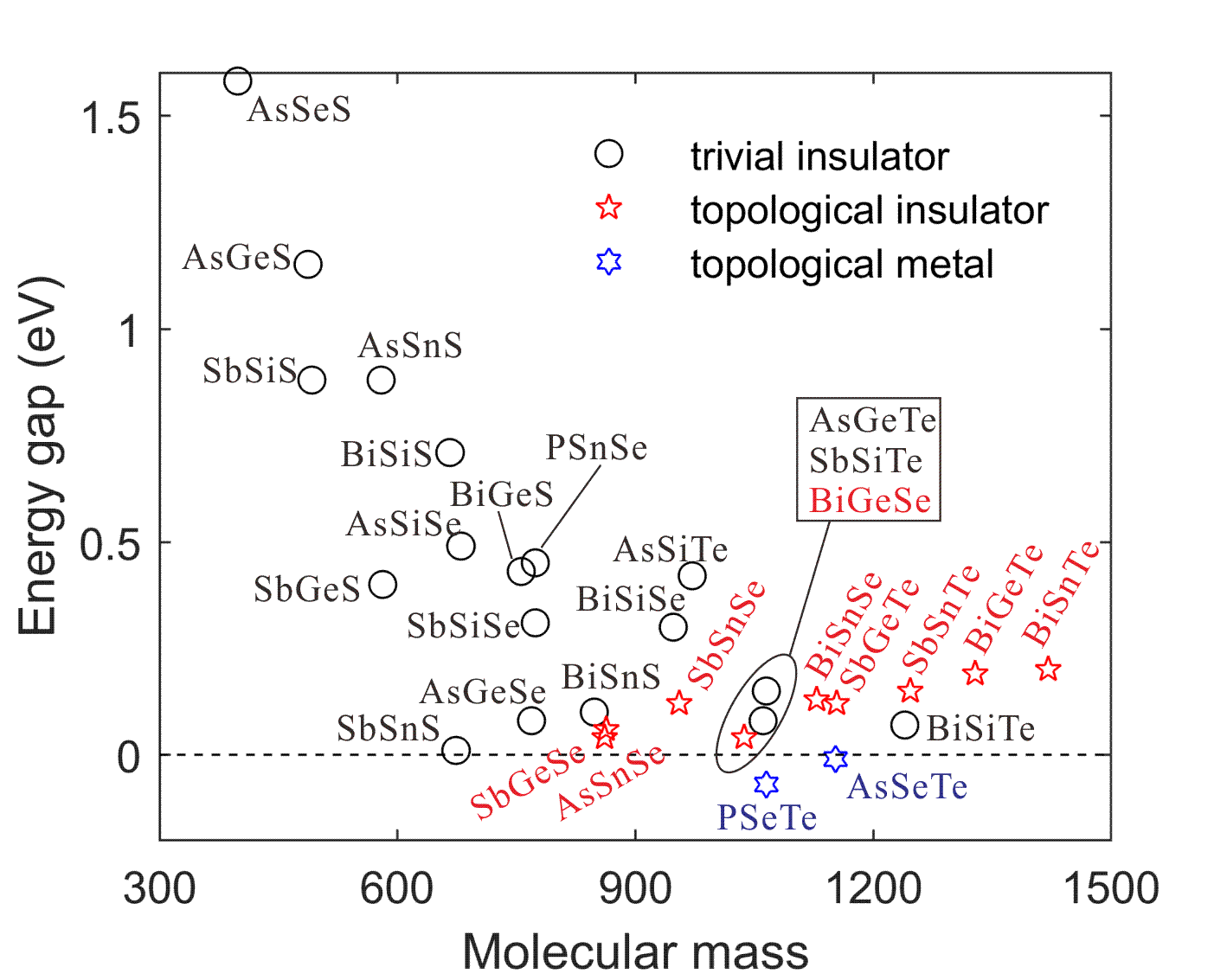}
\caption{Topological nature and energy gaps of various V$_2$IV$_2$VI$_6$ materials as a function of molecular mass. The subscripts in the molecular formulas have been neglected for clarity. }
\label{phase}
\end{figure*}

Band structure calculations indicate that these stable materials are either insulators or such metals with local energy gaps everywhere around Fermi level in the Brillouin zone. Then $Z_2$ topological invariants ($\nu_0; \nu_1\nu_2\nu_3$) \cite{prl106803, prb121306, prb195322} are calculated using a parity method proposed by Fu and Kane \cite{prb045302}, according to which the materials are classified into three topological classes, i.e., TIs, TMs and trivial insulators. The topological phases and energy gaps of these materials are shown in Fig.~\ref{phase}, where the subscripts in the molecular formulas are neglected for clarity. The lattice constants, gap values and $Z_2$ topological invariants of TIs and TMs are listed in Table~\ref{table1}, while those of the trivial insulators are listed in Table ~\ref{table2}.

\begin{table*}[htb]
\caption{Theoretical parameters of nontrivial V$_2$IV$_2$VI$_6$ crystals: lattice constants \textit{a} and \textit{c}, dynamically stable (Y) or unstable (N), energy gap E$_g$, and topological invariants ($\nu_0; \nu_1\nu_2\nu_3$). Minus gaps of As$_2$Sn$_2$Te$_6$ and P$_2$Sn$_2$Te$_6$ mean topological metals. The superscripts \textit{d} and \textit{i} on the gap data indicate the direct and indirect gap nature, respectively. }
\centering  
\begin{tabular}{L{2.cm}C{1.6cm}C{1.6cm}C{1.5cm}C{1.5cm}C{2.2cm}} 
\hline \hline
 &\textit{a} (\r{A}) &\textit{c} (\r{A}) &Stability &E$_g$ (eV) &$Z_2$ class \\ \hline  
As$_2$Sn$_2$Se$_6$ &6.757 &20.011 &Y &0.04$^i$ &(1;111)\\

Sb$_2$Ge$_2$Se$_6$ &6.816 &19.949 &Y &0.06$^i$ &(1;111)\\
Sb$_2$Ge$_2$Te$_6$ &7.302 &21.010 &Y &0.12$^i$ &(1;111)\\
Sb$_2$Sn$_2$Se$_6$ &6.974 &20.153 &Y &0.12$^i$ &(1;111)\\
Sb$_2$Sn$_2$Te$_6$ &7.433 &21.467 &Y &0.15$^i$ &(1;111)\\

Bi$_2$Ge$_2$Se$_6$ &6.909 &20.197 &Y &0.04$^d$ &(1;111)\\
Bi$_2$Ge$_2$Te$_6$ &7.392 &21.159 &Y &0.19$^i$ &(1;111)\\
Bi$_2$Sn$_2$Se$_6$ &7.064 &20.334 &Y &0.13$^i$ &(1;111)\\
Bi$_2$Sn$_2$Te$_6$ &7.521 &21.637 &Y &0.20$^i$ &(1;111)\\
P$_2$Sn$_2$Te$_6$ &7.147 &20.880 &Y &$-0.07$ &(1;111)\\
As$_2$Sn$_2$Te$_6$ &7.223 &21.092 &Y &$-0.01$ &(1;111)\\\hline
\end{tabular}
\label{table1}
\end{table*}

\begin{table*}[tbh]
\caption{Theoretical parameters of trivial V$_2$IV$_2$VI$_6$ crystals: lattice constants \textit{a} and \textit{c}, dynamically stable (Y) or unstable (N), energy gap E$_g$, and topological Z$_2$ invariants. The superscripts \textit{d} and \textit{i} on the gap data indicate the direct and indirect gap nature, respectively. }
\centering  
\begin{tabular}{L{2.cm}C{1.6cm}C{1.6cm}C{2.cm}C{2.0cm}C{2.2cm}} 
\hline \hline
 &\textit{a} (\r{A}) &\textit{c} (\r{A}) &Stability &E$_g$ (eV) &Z$_2$ invariants\\ \hline  
P$_2$Sn$_2$Se$_6$ &6.543 &23.491 &Y &0.45$^d$ &0; (000)\\
As$_2$Si$_2$S$_6$ &6.124 &20.613 &Y &1.58$^i$ &0; (000)\\
As$_2$Si$_2$Se$_6$ &6.467 &19.902 &Y &0.49$^d$ &0; (000)\\
As$_2$Si$_2$Te$_6$ &6.986 &20.931 &Y &0.42$^d$ &0; (000)\\
As$_2$Ge$_2$S$_6$ &6.209 &21.035 &Y &1.15$^d$ &0; (000)\\
As$_2$Ge$_2$Se$_6$ &6.564 &19.779 &Y &0.08$^i$ &0; (000)\\
As$_2$Ge$_2$Te$_6$ &7.069 &20.774 &Y &0.08$^d$ &0; (000)\\
As$_2$Sn$_2$S$_6$ &6.370 &22.051 &Y &0.88$^d$ &0; (000)\\
Sb$_2$Si$_2$S$_6$ &6.428 &19.454 &Y &0.88$^d$ &0; (000)\\
Sb$_2$Si$_2$Se$_6$ &6.722 &20.155 &Y &0.31$^d$ &0; (000)\\
Sb$_2$Si$_2$Te$_6$ &7.223 &21.236 &Y &0.15$^d$ &0; (000)\\
Sb$_2$Ge$_2$S$_6$ &6.534 &19.378 &Y &0.40$^d$ &0; (000)\\
Sb$_2$Sn$_2$S$_6$ &6.721 &19.298 &Y &0.01$^d$ &0; (000)\\
Bi$_2$Si$_2$S$_6$ &6.537 &19.608 &Y &0.71$^d$ &0; (000)\\
Bi$_2$Si$_2$Se$_6$ &6.821 &20.419 &Y &0.30$^d$ &0; (000)\\
Bi$_2$Si$_2$Te$_6$ &7.316 &21.438 &Y &0.07$^d$ &0; (000)\\
Bi$_2$Ge$_2$S$_6$ &6.630 &19.536 &Y &0.43$^d$ &0; (000)\\
Bi$_2$Sn$_2$S$_6$ &6.807 &19.499 &Y &0.10$^d$ &0; (000)\\
Bi$_2$Pb$_2$Se$_6$ &- &- &N &-  &-\\
Bi$_2$Pb$_2$Te$_6$ &- &- &N &-  &-\\ \hline
\end{tabular}
\label{table2}
\end{table*}

Fig.~\ref{phase} shows two metals, P$_2$Sn$_2$Te$_6$ and As$_2$Sn$_2$Te$_6$. Band structure of P$_2$Sn$_2$Te$_6$ is shown in Fig.~\ref{band2} (d) as an example, and that of As$_2$Sn$_2$Te$_6$ is similar. The CBM of P$_2$Sn$_2$Te$_6$ is located at F point, about 0.07 eV lower than the VBM, which is close to T point. The compound is a metal. However, local energy gaps exist around the Fermi level everywhere in the Brillouin zone, rendering the computation of topological invariants possible. According to the calculation results listed in Table~\ref{table1}, both P$_2$Sn$_2$Te$_6$ and As$_2$Sn$_2$Te$_6$ are in the nontrivial (1;111) class, thus they are strong TMs. Recently, multiple Dirac cones have been reported in Zr$_2$Te$_2$P \cite{prb045315}, which is a 3D strong TM that have the same symmetry group as P$_2$Sn$_2$Te$_6$ and As$_2$Sn$_2$Te$_6$.

Fig.~\ref{phase} also shows nine TIs, including As$_2$Sn$_2$Se$_6$ and eight V$_2$IV$_2$VI$_6$ crystals with V=Sb, Bi, IV=Ge, Sn, VI=Se, Te. These TIs have narrow energy gaps ranging from 0.04 to 0.20 eV and include heavy elements with strong SOC, which are usually essential for TIs \cite{rmp3045, rmp1057}. However, not all the materials containing heavy elements are TIs. Sb$_2$Si$_2$Te$_6$, Bi$_2$Si$_2$Te$_6$ and Bi$_2$Sn$_2$S$_6$ are trivial insulators due to the involvement of the light Si or S atoms. As$_2$Ge$_2$Se$_6$ is also topologically trivial because both As and Ge are not heavy enough. From Table~\ref{table1}, these TIs are all in the (1;111) $Z_2$ class, indicating they are 3D strong TIs. As comparisons, the Bi$_{1-x}$Sb$_x$ alloy and Bi$_2$Se$_3$ are 3D strong TIs in the (1,111) and (1;000) topological classes \cite{prb045426, cpc1849}, respectively.

The inplane lattice constants of the TIs range from about 6.7 to 7.5 \r{A}, covering the lattice regions of Cr$_2$Ge$_2$Te$_6$ and CrI$_3$, which are typically around 7.0 \r{A} \cite{arxiv07358, arxiv02560}. Cr$_2$Ge$_2$Te$_6$ \cite{nature47} and CrI$_3$ \cite{science1214, science1218, nature270} are vdW-layered ferromagnetic insulators (FMIs) that have attracted very much attention recently. FMI/TI heterostructures such as Cr$_2$Ge$_2$Te$_6$/Bi$_2$Te$_3$ \cite{jap114907}, Cr$_2$Ge$_2$Te$_6$/BiSbTeSe$_2$ \cite{nl8047} and CrI$_3$/Bi$_2$Se$_3$ \cite{arxiv07358} have been investigated to realize exotic quantum phenomena such as quantum anomalous Hall effect and so on. Compared with the Bi$_2$Se$_3$ family of TIs, whose inplane lattice constants vary from 4.1 to 4.3 \r{A} \cite{njp065013}, the current TIs match Cr$_2$Ge$_2$Te$_6$ and CrI$_3$ well. Hence, it is interesting to study the magnetic proximity effects between these FMIs and current TIs.

\begin{figure*}[thp]
\includegraphics [width=12cm] {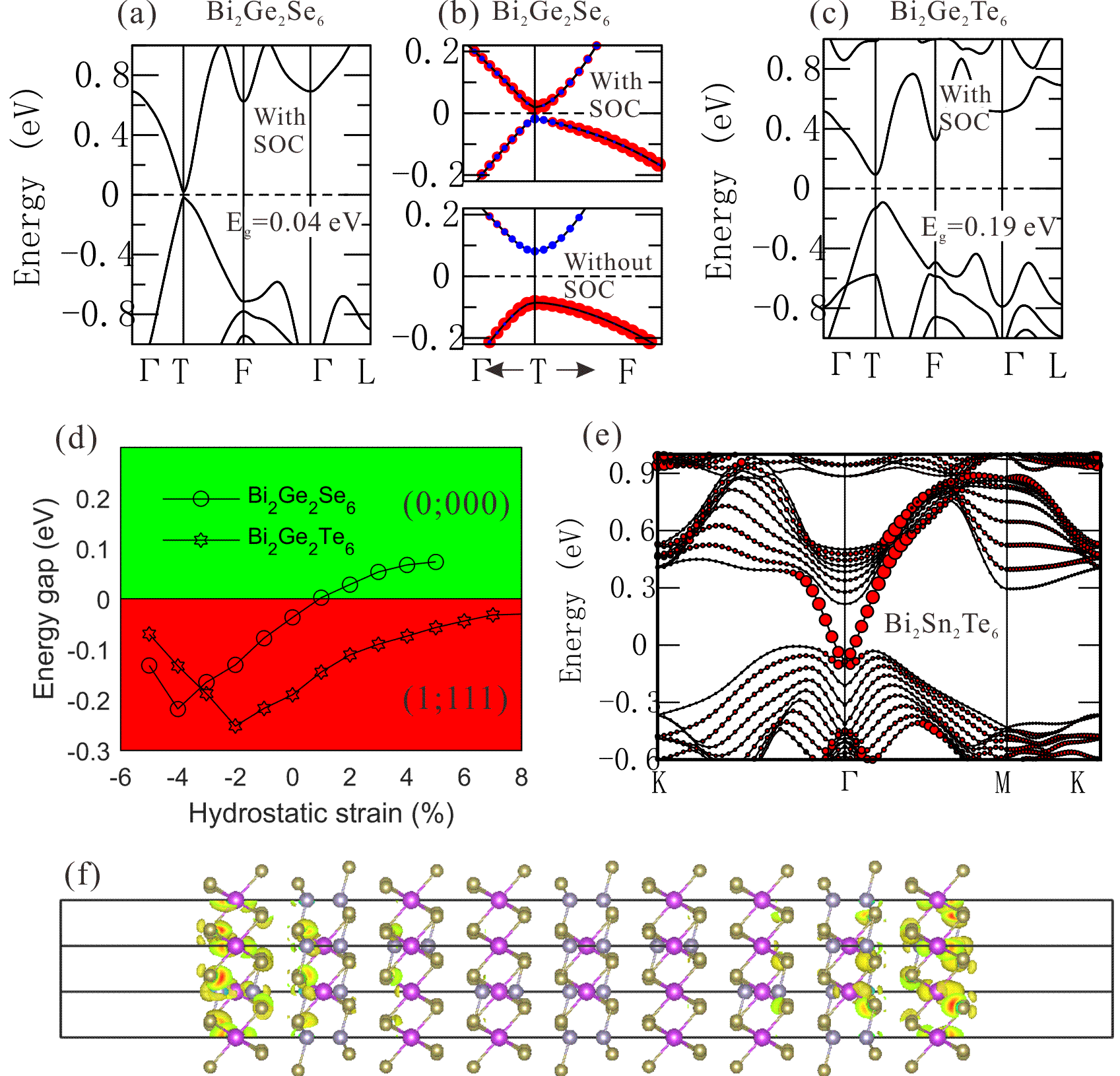}
\caption{(a) Band structure and (b) enlarged band structure around T point of Bi$_2$Ge$_2$Se$_6$. Red and blue symbols indicate the Se-p and the Bi-p states, respectively. (c) Band structure of Bi$_2$Ge$_2$Te$_6$. (d) Energy gaps at different hydrostatic strains of Bi$_2$Ge$_2$Se$_6$ and Bi$_2$Ge$_2$Te$_6$. (e) Band structure of the Bi$_2$Sn$_2$Te$_6$ slab, where the red symbols denote contributions of surface atoms. (f) Projected wavefunctions of the Dirac point of the Bi$_2$Sn$_2$Te$_6$ slab. }
\label{band1}
\end{figure*}

Among the nine TIs, only Bi$_2$Ge$_2$Se$_6$ has a direct energy gap. Its band structure, and zoomed-in band structures are shown in Fig.~\ref{band1} (a) and (b), respectively. Both the VBM and the CBM are located at T point in the reciprocal space, forming a direct gap of 0.04 eV. When spin-orbit coupling (SOC) is considered, the VBM and the CBM of Bi$_2$Ge$_2$Se$_6$ are mainly contributed by the Bi-p and the Te-p states, respectively. If SOC is turned off, the VBM and the CBM mainly come from the Te-p and the Bi-p states, respectively. Hence, SOC results in the band inversion, which is a common mechanism of TIs \cite{rmp3045, rmp1057}.

Other TIs are indirectly gapped. As$_2$Sn$_2$Se$_6$ has the smallest energy gap of 0.04 eV, equal to that of Bi$_2$Ge$_2$Se$_6$. Bi$_2$Ge$_2$Te$_6$ and Bi$_2$Sn$_2$Te$_6$ have the largest energy gaps, 0.19 and 0.20 eV, respectively, in the same order as those of the Bi$_2$Se$_3$ family of TIs \cite{np438}, and large enough for room temperature utilizations. Consider, for example, the case of Bi$_2$Ge$_2$Te$_6$. Its CBM is located at T point, but the VBM shifts away along the TF direction, as shown in Fig.~\ref{band1} (c), forming an indirect energy gap. When not considering SOC, the CBM and the VBM of Bi$_2$Ge$_2$Te$_6$ mainly come from the Te-p and the Bi-p states. When considering SOC, the energy bands near the Fermi level are inverted around T point, very similar to the case of Bi$_2$Ge$_2$Se$_6$.

Strains can significantly change the electronic states of the current TIs. We discuss two examples, Bi$_2$Ge$_2$Se$_6$ and Bi$_2$Ge$_2$Te$_6$. Fig.~\ref{band1} (d) shows the gap-dependence on the hydrostatic strain $\varepsilon$. In both cases, small compressive strains remarkably increase the energy gaps. For example, a 4\% compressive strain increases the energy gap of Bi$_2$Ge$_2$Se$_6$ from 0.04 to 0.22 eV. However, after that, the gap drastically decreases as the compressive strain increases. Similar behavior also occurs in Bi$_2$Ge$_2$Te$_6$ after a compressive strain of 2\%. This is because after the critical strains of 4\% and 2\%, the CBMs of the both materials shift from T to F point in the reciprocal space. If the compressive strain continues, energy levels of the new CBMs decrease more quickly with respect to their respective VBMs.

Under tensile strains, Bi$_2$Ge$_2$Se$_6$ and Bi$_2$Ge$_2$Te$_6$ behave differently. At $\varepsilon<1\%$, Bi$_2$Ge$_2$Se$_6$ is gapped, corresponding to a nontrivial phase or a TI. At $\varepsilon=1\%$, the VBM and the CBM of Bi$_2$Ge$_2$Se$_6$ touch at T point, so the gap closes, forming a semimetal. At $\varepsilon>1\%$, the gap reopens, producing an insulator. This gap closing and reopening strongly suggests the transition between nontrivial and trivial phases \cite{njp065013}. We confirm this by computing the $Z_2$ invariants, which are (1;111) and (0;000) before and after the gap reopening, respectively. Thus, after the gap reopens, Bi$_2$Ge$_2$Se$_6$ becomes a trivial insulator. At $\varepsilon=2\%$, the energy gap is about 0.03 eV, which increases to 0.07 eV at $\varepsilon=5\%$. In the case of Bi$_2$Ge$_2$Te$_6$, the energy gap decreases with the increasing of the tensile strain, but that slows down as the strain becomes large. At $\varepsilon=8\%$, the energy gap is still 0.03 eV. Further analyses indicate that gap remains nonzero until the material becomes metallic at an extremely large tensile strain of 14\%. In this process, there are no gap closing and reopening, thus no topological phase transitions. It illustrates the robustness of a TI against nonmagnetic perturbations like strains \cite{rmp3045, rmp1057}.

TIs distinguish from trivial insulators by their massless and linearly dispersed Dirac surface states, which are protected by time-reversal symmetry  \cite{rmp3045, rmp1057}. The predicted TIs support a single Dirac cone on the surface. Take Bi$_2$Sn$_2$Te$_6$ as an example. Fig.~\ref{band1} (e) shows the band structure of a Bi$_2$Sn$_2$Te$_6$ slab of nine  QLs, where the surface contribution is indicated by red symbols. A single Dirac cone is easily resolved, with the Dirac point at $\Gamma$ point, and it obviously comes from surface atoms. The projected wavefunction of the Dirac point in the real space shows that the surface states penetrate a depth of about two QLs, as demonstrated in Fig.~\ref{band1} (f). Band structure of more TI slabs are shown in Fig.~\ref{ext1-sslab} in Supplementary Materials.

Besides the above TMs and TIs, Fig.~\ref{phase} shows 18 trivial insulators in the topological class (0;000), some parameters of which are listed in Table ~\ref{table2}. They involve light atoms such as S, Si and P with weak SOC. Most of them are directly gapped. Both the CBM and the VBM are located at T point, as the example Bi$_2$Si$_2$Te$_6$ shows in Fig.~\ref{band2} (a). As$_2$Si$_2$S$_6$ and As$_2$Ge$_2$Se$_6$ are indirectly gapped. In the As$_2$Si$_2$S$_6$ case, the CBM and the VBM are at L and T points, respectively. In the As$_2$Ge$_2$Se$_6$ case, the CBM is also at T point, but the VBM slightly shifts away from T point along the TF direction. The lattice constants of these trivial insulators vary from about 6.1 to 7.3 \r{A}, and their energy gaps range from 0.01 to 1.58 eV. The smallest and largest energy gaps occur in Sb$_2$Sn$_2$S$_6$ and As$_2$Si$_2$S$_6$, respectively.

Topological phase transitions can be easily controlled by applied strains in some trivial insulators. Take Sb$_2$Sn$_2$S$_6$ as an example. At equilibrium lattice constants, Sb$_2$Sn$_2$S$_6$ is trivial, with a small direct energy gap 0.01 eV. However, as Fig.~\ref{band2} (b) shows it becomes nontrivial at $\varepsilon=-1\%$ according to the calculated $Z_2$ invariants.

\section{Discussion}
The studied materials are all vdW-layered, periodically stacked along the [111] direction by V$_2$IV$_2$VI$_6$ QLs. Epitaxially grown thin films of such materials can mostly avoid defective surface defects, and slabs or films can also be exfoliated from their 3D bulk. Phonon dispersions (see Fig.~\ref{ext1-sslab} in Supplementary Materials) show that slabs of these materials are dynamically stable down to one QL. These TIs have large energy gaps enough for room temperature utilizations. Their lattice constants vary from 6.7 to 7.5 \r{A}, covering the lattice range of Cr$_2$Ge$_2$Te$_6$ and CrI$_3$ \cite{arxiv07358, arxiv02560}. Hence, they are promising candidates to study the topological-magnetic proximity effect with Cr$_2$Ge$_2$Te$_6$ and CrI$_3$. Under small compressive strains, more trivial insulators can be turned into TIs. Several compounds in the family have been successfully synthesized in experiment \cite{dis2010, dis2014}, thus the synthesis of other materials of this family are quite promising and worth trying.

In conclusion, a number of 3D strong TIs and TMs are predicted based on first-principles calculations. They are vdW-layered, exhibit large energy gaps and wide range of lattice constants. This work greatly enriches the family of TIs, and provides ideal candidates for the study of topological-magnetic interactions with ferromagnetic insulators Cr$_2$Ge$_2$Te$_6$ and CrI$_3$. Synthesis of these materials is recommended to follow the successful approach synthesizing Bi$_2$Si$_2$Te$_6$ and Sb$_2$Si$_2$Te$_6$. We call for such efforts.

This work is supported by the NSF of China (Grant Nos. 21271007 and 11747016), the NSF of Anhui Province (Grant No. 1308085QA05) and the Doctoral Foundation of Anhui Jianzhu University (Grant No. 2017QD19).

\bibliography{refs}

\begin{thebibliography}{39}%
\makeatletter
\providecommand \@ifxundefined [1]{%
 \@ifx{#1\undefined}
}%
\providecommand \@ifnum [1]{%
 \ifnum #1\expandafter \@firstoftwo
 \else \expandafter \@secondoftwo
 \fi
}%
\providecommand \@ifx [1]{%
 \ifx #1\expandafter \@firstoftwo
 \else \expandafter \@secondoftwo
 \fi
}%
\providecommand \natexlab [1]{#1}%
\providecommand \enquote  [1]{``#1''}%
\providecommand \bibnamefont  [1]{#1}%
\providecommand \bibfnamefont [1]{#1}%
\providecommand \citenamefont [1]{#1}%
\providecommand \href@noop [0]{\@secondoftwo}%
\providecommand \href [0]{\begingroup \@sanitize@url \@href}%
\providecommand \@href[1]{\@@startlink{#1}\@@href}%
\providecommand \@@href[1]{\endgroup#1\@@endlink}%
\providecommand \@sanitize@url [0]{\catcode `\\12\catcode `\$12\catcode
  `\&12\catcode `\#12\catcode `\^12\catcode `\_12\catcode `\%12\relax}%
\providecommand \@@startlink[1]{}%
\providecommand \@@endlink[0]{}%
\providecommand \url  [0]{\begingroup\@sanitize@url \@url }%
\providecommand \@url [1]{\endgroup\@href {#1}{\urlprefix }}%
\providecommand \urlprefix  [0]{URL }%
\providecommand \Eprint [0]{\href }%
\providecommand \doibase [0]{http://dx.doi.org/}%
\providecommand \selectlanguage [0]{\@gobble}%
\providecommand \bibinfo  [0]{\@secondoftwo}%
\providecommand \bibfield  [0]{\@secondoftwo}%
\providecommand \translation [1]{[#1]}%
\providecommand \BibitemOpen [0]{}%
\providecommand \bibitemStop [0]{}%
\providecommand \bibitemNoStop [0]{.\EOS\space}%
\providecommand \EOS [0]{\spacefactor3000\relax}%
\providecommand \BibitemShut  [1]{\csname bibitem#1\endcsname}%
\let\auto@bib@innerbib\@empty
\bibitem [{\citenamefont {Fu}\ \emph {et~al.}(2007)\citenamefont {Fu},
  \citenamefont {Kane},\ and\ \citenamefont {Mele}}]{prl106803}%
  \BibitemOpen
  \bibfield  {author} {\bibinfo {author} {\bibfnamefont {L.}~\bibnamefont
  {Fu}}, \bibinfo {author} {\bibfnamefont {C.~L.}\ \bibnamefont {Kane}}, \ and\
  \bibinfo {author} {\bibfnamefont {E.~J.}\ \bibnamefont {Mele}},\ }\href
  {\doibase 10.1103/PhysRevLett.98.106803} {\bibfield  {journal} {\bibinfo
  {journal} {Phys. Rev. Lett.}\ }\textbf {\bibinfo {volume} {98}},\ \bibinfo
  {pages} {106803} (\bibinfo {year} {2007})}\BibitemShut {NoStop}%
\bibitem [{\citenamefont {Moore}\ and\ \citenamefont
  {Balents}(2007)}]{prb121306}%
  \BibitemOpen
  \bibfield  {author} {\bibinfo {author} {\bibfnamefont {J.~E.}\ \bibnamefont
  {Moore}}\ and\ \bibinfo {author} {\bibfnamefont {L.}~\bibnamefont
  {Balents}},\ }\href {\doibase 10.1103/PhysRevB.75.121306} {\bibfield
  {journal} {\bibinfo  {journal} {Phys. Rev. B}\ }\textbf {\bibinfo {volume}
  {75}},\ \bibinfo {pages} {121306(R)} (\bibinfo {year} {2007})}\BibitemShut
  {NoStop}%
\bibitem [{\citenamefont {Roy}(2009)}]{prb195322}%
  \BibitemOpen
  \bibfield  {author} {\bibinfo {author} {\bibfnamefont {R.}~\bibnamefont
  {Roy}},\ }\href {\doibase 10.1103/PhysRevB.79.195322} {\bibfield  {journal}
  {\bibinfo  {journal} {Phys. Rev. B}\ }\textbf {\bibinfo {volume} {79}},\
  \bibinfo {pages} {195322} (\bibinfo {year} {2009})}\BibitemShut {NoStop}%
\bibitem [{\citenamefont {Fu}\ and\ \citenamefont {Kane}(2007)}]{prb045302}%
  \BibitemOpen
  \bibfield  {author} {\bibinfo {author} {\bibfnamefont {L.}~\bibnamefont
  {Fu}}\ and\ \bibinfo {author} {\bibfnamefont {C.~L.}\ \bibnamefont {Kane}},\
  }\href {\doibase 10.1103/PhysRevB.76.045302} {\bibfield  {journal} {\bibinfo
  {journal} {Phys. Rev. B}\ }\textbf {\bibinfo {volume} {76}},\ \bibinfo
  {pages} {045302} (\bibinfo {year} {2007})}\BibitemShut {NoStop}%
\bibitem [{\citenamefont {Xia}\ \emph {et~al.}(2009)\citenamefont {Xia},
  \citenamefont {Qian}, \citenamefont {Hsieh}, \citenamefont {Wray},
  \citenamefont {Pal}, \citenamefont {Lin}, \citenamefont {Bansil},
  \citenamefont {Grauer}, \citenamefont {Hor}, \citenamefont {Cava},\ and\
  \citenamefont {Hasan}}]{np398}%
  \BibitemOpen
  \bibfield  {author} {\bibinfo {author} {\bibfnamefont {Y.}~\bibnamefont
  {Xia}}, \bibinfo {author} {\bibfnamefont {D.}~\bibnamefont {Qian}}, \bibinfo
  {author} {\bibfnamefont {D.}~\bibnamefont {Hsieh}}, \bibinfo {author}
  {\bibfnamefont {L.}~\bibnamefont {Wray}}, \bibinfo {author} {\bibfnamefont
  {A.}~\bibnamefont {Pal}}, \bibinfo {author} {\bibfnamefont {H.}~\bibnamefont
  {Lin}}, \bibinfo {author} {\bibfnamefont {A.}~\bibnamefont {Bansil}},
  \bibinfo {author} {\bibfnamefont {D.}~\bibnamefont {Grauer}}, \bibinfo
  {author} {\bibfnamefont {Y.~S.}\ \bibnamefont {Hor}}, \bibinfo {author}
  {\bibfnamefont {R.~J.}\ \bibnamefont {Cava}}, \ and\ \bibinfo {author}
  {\bibfnamefont {M.~Z.}\ \bibnamefont {Hasan}},\ }\href {\doibase
  10.1038/nphys1274} {\bibfield  {journal} {\bibinfo  {journal} {Nat. Phys.}\
  }\textbf {\bibinfo {volume} {5}},\ \bibinfo {pages} {398} (\bibinfo {year}
  {2009})}\BibitemShut {NoStop}%
\bibitem [{\citenamefont {Zhang}\ \emph {et~al.}(2009)\citenamefont {Zhang},
  \citenamefont {Liu}, \citenamefont {Qi}, \citenamefont {Dai}, \citenamefont
  {Fang},\ and\ \citenamefont {Zhang}}]{np438}%
  \BibitemOpen
  \bibfield  {author} {\bibinfo {author} {\bibfnamefont {H.}~\bibnamefont
  {Zhang}}, \bibinfo {author} {\bibfnamefont {C.-X.}\ \bibnamefont {Liu}},
  \bibinfo {author} {\bibfnamefont {X.-L.}\ \bibnamefont {Qi}}, \bibinfo
  {author} {\bibfnamefont {X.}~\bibnamefont {Dai}}, \bibinfo {author}
  {\bibfnamefont {Z.}~\bibnamefont {Fang}}, \ and\ \bibinfo {author}
  {\bibfnamefont {S.-C.}\ \bibnamefont {Zhang}},\ }\href {\doibase
  10.1038/nphys1270} {\bibfield  {journal} {\bibinfo  {journal} {Nat. Phys.}\
  }\textbf {\bibinfo {volume} {5}},\ \bibinfo {pages} {438} (\bibinfo {year}
  {2009})}\BibitemShut {NoStop}%
\bibitem [{\citenamefont {Hasan}\ and\ \citenamefont {Kane}(2010)}]{rmp3045}%
  \BibitemOpen
  \bibfield  {author} {\bibinfo {author} {\bibfnamefont {M.~Z.}\ \bibnamefont
  {Hasan}}\ and\ \bibinfo {author} {\bibfnamefont {C.~L.}\ \bibnamefont
  {Kane}},\ }\href {\doibase 10.1103/RevModPhys.82.3045} {\bibfield  {journal}
  {\bibinfo  {journal} {Rev. Mod. Phys.}\ }\textbf {\bibinfo {volume} {82}},\
  \bibinfo {pages} {3045} (\bibinfo {year} {2010})}\BibitemShut {NoStop}%
\bibitem [{\citenamefont {Qi}\ and\ \citenamefont {Zhang}(2011)}]{rmp1057}%
  \BibitemOpen
  \bibfield  {author} {\bibinfo {author} {\bibfnamefont {X.-L.}\ \bibnamefont
  {Qi}}\ and\ \bibinfo {author} {\bibfnamefont {S.-C.}\ \bibnamefont {Zhang}},\
  }\href {\doibase 10.1103/RevModPhys.83.1057} {\bibfield  {journal} {\bibinfo
  {journal} {Rev. Mod. Phys.}\ }\textbf {\bibinfo {volume} {83}},\ \bibinfo
  {pages} {1057} (\bibinfo {year} {2011})}\BibitemShut {NoStop}%
\bibitem [{\citenamefont {Agarwala}\ and\ \citenamefont
  {Shenoy}(2017)}]{prl236402}%
  \BibitemOpen
  \bibfield  {author} {\bibinfo {author} {\bibfnamefont {A.}~\bibnamefont
  {Agarwala}}\ and\ \bibinfo {author} {\bibfnamefont {V.~B.}\ \bibnamefont
  {Shenoy}},\ }\href {\doibase 10.1103/PhysRevLett.118.236402} {\bibfield
  {journal} {\bibinfo  {journal} {Phys. Rev. Lett.}\ }\textbf {\bibinfo
  {volume} {118}},\ \bibinfo {pages} {236402} (\bibinfo {year}
  {2017})}\BibitemShut {NoStop}%
\bibitem [{\citenamefont {Pi}\ \emph {et~al.}(2017)\citenamefont {Pi},
  \citenamefont {Wang}, \citenamefont {Kim}, \citenamefont {Wu}, \citenamefont
  {Wang},\ and\ \citenamefont {Lu}}]{jpcl332}%
  \BibitemOpen
  \bibfield  {author} {\bibinfo {author} {\bibfnamefont {S.-T.}\ \bibnamefont
  {Pi}}, \bibinfo {author} {\bibfnamefont {H.}~\bibnamefont {Wang}}, \bibinfo
  {author} {\bibfnamefont {J.}~\bibnamefont {Kim}}, \bibinfo {author}
  {\bibfnamefont {R.}~\bibnamefont {Wu}}, \bibinfo {author} {\bibfnamefont
  {Y.-K.}\ \bibnamefont {Wang}}, \ and\ \bibinfo {author} {\bibfnamefont
  {C.-K.}\ \bibnamefont {Lu}},\ }\href {\doibase 10.1021/acs.jpclett.6b02860}
  {\bibfield  {journal} {\bibinfo  {journal} {J. Phys. Chem. Lett.}\ }\textbf
  {\bibinfo {volume} {8}},\ \bibinfo {pages} {332} (\bibinfo {year}
  {2017})}\BibitemShut {NoStop}%
\bibitem [{\citenamefont {Pielnhofer}\ \emph {et~al.}(2017)\citenamefont
  {Pielnhofer}, \citenamefont {Menshchikova}, \citenamefont {Rusinov},
  \citenamefont {Zeugner}, \citenamefont {Sklyadneva}, \citenamefont {Heid},
  \citenamefont {Bohnen}, \citenamefont {Golub}, \citenamefont {Baranov},
  \citenamefont {Chulkov}, \citenamefont {Pfitzner}, \citenamefont {Ruck},\
  and\ \citenamefont {Isaeva}}]{jmcc4752}%
  \BibitemOpen
  \bibfield  {author} {\bibinfo {author} {\bibfnamefont {F.}~\bibnamefont
  {Pielnhofer}}, \bibinfo {author} {\bibfnamefont {T.~V.}\ \bibnamefont
  {Menshchikova}}, \bibinfo {author} {\bibfnamefont {I.~P.}\ \bibnamefont
  {Rusinov}}, \bibinfo {author} {\bibfnamefont {A.}~\bibnamefont {Zeugner}},
  \bibinfo {author} {\bibfnamefont {I.~Y.}\ \bibnamefont {Sklyadneva}},
  \bibinfo {author} {\bibfnamefont {R.}~\bibnamefont {Heid}}, \bibinfo {author}
  {\bibfnamefont {K.-P.}\ \bibnamefont {Bohnen}}, \bibinfo {author}
  {\bibfnamefont {P.}~\bibnamefont {Golub}}, \bibinfo {author} {\bibfnamefont
  {A.~I.}\ \bibnamefont {Baranov}}, \bibinfo {author} {\bibfnamefont {E.~V.}\
  \bibnamefont {Chulkov}}, \bibinfo {author} {\bibfnamefont {A.}~\bibnamefont
  {Pfitzner}}, \bibinfo {author} {\bibfnamefont {M.}~\bibnamefont {Ruck}}, \
  and\ \bibinfo {author} {\bibfnamefont {A.}~\bibnamefont {Isaeva}},\ }\href
  {\doibase 10.1039/C7TC00390K} {\bibfield  {journal} {\bibinfo  {journal} {J.
  Mater. Chem. C}\ }\textbf {\bibinfo {volume} {5}},\ \bibinfo {pages} {4752}
  (\bibinfo {year} {2017})}\BibitemShut {NoStop}%
\bibitem [{\citenamefont {Young}\ \emph {et~al.}(2017)\citenamefont {Young},
  \citenamefont {Manni}, \citenamefont {Shao}, \citenamefont {Canfield},\ and\
  \citenamefont {Kolmogorov}}]{prb085116}%
  \BibitemOpen
  \bibfield  {author} {\bibinfo {author} {\bibfnamefont {S.~M.}\ \bibnamefont
  {Young}}, \bibinfo {author} {\bibfnamefont {S.}~\bibnamefont {Manni}},
  \bibinfo {author} {\bibfnamefont {J.}~\bibnamefont {Shao}}, \bibinfo {author}
  {\bibfnamefont {P.~C.}\ \bibnamefont {Canfield}}, \ and\ \bibinfo {author}
  {\bibfnamefont {A.~N.}\ \bibnamefont {Kolmogorov}},\ }\href {\doibase
  10.1103/PhysRevB.95.085116} {\bibfield  {journal} {\bibinfo  {journal} {Phys.
  Rev. B}\ }\textbf {\bibinfo {volume} {95}},\ \bibinfo {pages} {085116}
  (\bibinfo {year} {2017})}\BibitemShut {NoStop}%
\bibitem [{\citenamefont {Feng}\ and\ \citenamefont {Yao}(2012)}]{sc2199}%
  \BibitemOpen
  \bibfield  {author} {\bibinfo {author} {\bibfnamefont {W.}~\bibnamefont
  {Feng}}\ and\ \bibinfo {author} {\bibfnamefont {Y.}~\bibnamefont {Yao}},\
  }\href {\doibase 10.1007/s11433-012-4929-9} {\bibfield  {journal} {\bibinfo
  {journal} {Sci. China-Phys. Mech. Astron.}\ }\textbf {\bibinfo {volume}
  {55}},\ \bibinfo {pages} {2199} (\bibinfo {year} {2012})}\BibitemShut
  {NoStop}%
\bibitem [{\citenamefont {Lin}\ \emph {et~al.}(2010)\citenamefont {Lin},
  \citenamefont {Markiewicz}, \citenamefont {Wray}, \citenamefont {Fu},
  \citenamefont {Hasan},\ and\ \citenamefont {Bansil}}]{prl036404}%
  \BibitemOpen
  \bibfield  {author} {\bibinfo {author} {\bibfnamefont {H.}~\bibnamefont
  {Lin}}, \bibinfo {author} {\bibfnamefont {R.~S.}\ \bibnamefont {Markiewicz}},
  \bibinfo {author} {\bibfnamefont {L.~A.}\ \bibnamefont {Wray}}, \bibinfo
  {author} {\bibfnamefont {L.}~\bibnamefont {Fu}}, \bibinfo {author}
  {\bibfnamefont {M.~Z.}\ \bibnamefont {Hasan}}, \ and\ \bibinfo {author}
  {\bibfnamefont {A.}~\bibnamefont {Bansil}},\ }\href {\doibase
  10.1103/PhysRevLett.105.036404} {\bibfield  {journal} {\bibinfo  {journal}
  {Phys. Rev. Lett.}\ }\textbf {\bibinfo {volume} {105}},\ \bibinfo {pages}
  {036404} (\bibinfo {year} {2010})}\BibitemShut {NoStop}%
\bibitem [{\citenamefont {Kuroda}\ \emph {et~al.}(2010)\citenamefont {Kuroda},
  \citenamefont {Ye}, \citenamefont {Kimura}, \citenamefont {Eremeev},
  \citenamefont {Krasovskii}, \citenamefont {Chulkov}, \citenamefont {Ueda},
  \citenamefont {Miyamoto}, \citenamefont {Okuda}, \citenamefont {Shimada},
  \citenamefont {Namatame},\ and\ \citenamefont {Taniguchi}}]{prl146801}%
  \BibitemOpen
  \bibfield  {author} {\bibinfo {author} {\bibfnamefont {K.}~\bibnamefont
  {Kuroda}}, \bibinfo {author} {\bibfnamefont {M.}~\bibnamefont {Ye}}, \bibinfo
  {author} {\bibfnamefont {A.}~\bibnamefont {Kimura}}, \bibinfo {author}
  {\bibfnamefont {S.~V.}\ \bibnamefont {Eremeev}}, \bibinfo {author}
  {\bibfnamefont {E.~E.}\ \bibnamefont {Krasovskii}}, \bibinfo {author}
  {\bibfnamefont {E.~V.}\ \bibnamefont {Chulkov}}, \bibinfo {author}
  {\bibfnamefont {Y.}~\bibnamefont {Ueda}}, \bibinfo {author} {\bibfnamefont
  {K.}~\bibnamefont {Miyamoto}}, \bibinfo {author} {\bibfnamefont
  {T.}~\bibnamefont {Okuda}}, \bibinfo {author} {\bibfnamefont
  {K.}~\bibnamefont {Shimada}}, \bibinfo {author} {\bibfnamefont
  {H.}~\bibnamefont {Namatame}}, \ and\ \bibinfo {author} {\bibfnamefont
  {M.}~\bibnamefont {Taniguchi}},\ }\href {\doibase
  10.1103/PhysRevLett.105.146801} {\bibfield  {journal} {\bibinfo  {journal}
  {Phys. Rev. Lett.}\ }\textbf {\bibinfo {volume} {105}},\ \bibinfo {pages}
  {146801} (\bibinfo {year} {2010})}\BibitemShut {NoStop}%
\bibitem [{\citenamefont {Kuroda}\ \emph {et~al.}(2012)\citenamefont {Kuroda},
  \citenamefont {Miyahara}, \citenamefont {Ye}, \citenamefont {Eremeev},
  \citenamefont {Koroteev}, \citenamefont {Krasovskii}, \citenamefont
  {Chulkov}, \citenamefont {Hiramoto}, \citenamefont {Moriyoshi}, \citenamefont
  {Kuroiwa}, \citenamefont {Miyamoto}, \citenamefont {Okuda}, \citenamefont
  {Arita}, \citenamefont {Shimada}, \citenamefont {Namatame}, \citenamefont
  {Taniguchi}, \citenamefont {Ueda},\ and\ \citenamefont {Kimura}}]{prl206803}%
  \BibitemOpen
  \bibfield  {author} {\bibinfo {author} {\bibfnamefont {K.}~\bibnamefont
  {Kuroda}}, \bibinfo {author} {\bibfnamefont {H.}~\bibnamefont {Miyahara}},
  \bibinfo {author} {\bibfnamefont {M.}~\bibnamefont {Ye}}, \bibinfo {author}
  {\bibfnamefont {S.~V.}\ \bibnamefont {Eremeev}}, \bibinfo {author}
  {\bibfnamefont {Y.~M.}\ \bibnamefont {Koroteev}}, \bibinfo {author}
  {\bibfnamefont {E.~E.}\ \bibnamefont {Krasovskii}}, \bibinfo {author}
  {\bibfnamefont {E.~V.}\ \bibnamefont {Chulkov}}, \bibinfo {author}
  {\bibfnamefont {S.}~\bibnamefont {Hiramoto}}, \bibinfo {author}
  {\bibfnamefont {C.}~\bibnamefont {Moriyoshi}}, \bibinfo {author}
  {\bibfnamefont {Y.}~\bibnamefont {Kuroiwa}}, \bibinfo {author} {\bibfnamefont
  {K.}~\bibnamefont {Miyamoto}}, \bibinfo {author} {\bibfnamefont
  {T.}~\bibnamefont {Okuda}}, \bibinfo {author} {\bibfnamefont
  {M.}~\bibnamefont {Arita}}, \bibinfo {author} {\bibfnamefont
  {K.}~\bibnamefont {Shimada}}, \bibinfo {author} {\bibfnamefont
  {H.}~\bibnamefont {Namatame}}, \bibinfo {author} {\bibfnamefont
  {M.}~\bibnamefont {Taniguchi}}, \bibinfo {author} {\bibfnamefont
  {Y.}~\bibnamefont {Ueda}}, \ and\ \bibinfo {author} {\bibfnamefont
  {A.}~\bibnamefont {Kimura}},\ }\href {\doibase
  10.1103/PhysRevLett.108.206803} {\bibfield  {journal} {\bibinfo  {journal}
  {Phys. Rev. Lett.}\ }\textbf {\bibinfo {volume} {108}},\ \bibinfo {pages}
  {206803} (\bibinfo {year} {2012})}\BibitemShut {NoStop}%
\bibitem [{\citenamefont {Hsieh}\ \emph {et~al.}(2009)\citenamefont {Hsieh},
  \citenamefont {Xia}, \citenamefont {Qian}, \citenamefont {Wray},
  \citenamefont {Feier}, \citenamefont {Dil}, \citenamefont {Osterwalder},
  \citenamefont {Patthey}, \citenamefont {Fedorov}, \citenamefont {Lin},
  \citenamefont {Bansil}, \citenamefont {Grauer}, \citenamefont {Hor},
  \citenamefont {Cava},\ and\ \citenamefont {Hasan}}]{prl146401}%
  \BibitemOpen
  \bibfield  {author} {\bibinfo {author} {\bibfnamefont {D.}~\bibnamefont
  {Hsieh}}, \bibinfo {author} {\bibfnamefont {Y.}~\bibnamefont {Xia}}, \bibinfo
  {author} {\bibfnamefont {D.}~\bibnamefont {Qian}}, \bibinfo {author}
  {\bibfnamefont {L.}~\bibnamefont {Wray}}, \bibinfo {author} {\bibfnamefont
  {F.}~\bibnamefont {Feier}}, \bibinfo {author} {\bibfnamefont {J.~H.}\
  \bibnamefont {Dil}}, \bibinfo {author} {\bibfnamefont {J.}~\bibnamefont
  {Osterwalder}}, \bibinfo {author} {\bibfnamefont {L.}~\bibnamefont
  {Patthey}}, \bibinfo {author} {\bibfnamefont {A.~V.}\ \bibnamefont
  {Fedorov}}, \bibinfo {author} {\bibfnamefont {H.}~\bibnamefont {Lin}},
  \bibinfo {author} {\bibfnamefont {A.}~\bibnamefont {Bansil}}, \bibinfo
  {author} {\bibfnamefont {D.}~\bibnamefont {Grauer}}, \bibinfo {author}
  {\bibfnamefont {Y.~S.}\ \bibnamefont {Hor}}, \bibinfo {author} {\bibfnamefont
  {R.~J.}\ \bibnamefont {Cava}}, \ and\ \bibinfo {author} {\bibfnamefont
  {M.~Z.}\ \bibnamefont {Hasan}},\ }\href {\doibase
  10.1103/PhysRevLett.103.146401} {\bibfield  {journal} {\bibinfo  {journal}
  {Phys. Rev. Lett.}\ }\textbf {\bibinfo {volume} {103}},\ \bibinfo {pages}
  {146401} (\bibinfo {year} {2009})}\BibitemShut {NoStop}%
\bibitem [{\citenamefont {Chen}\ \emph {et~al.}(2009)\citenamefont {Chen},
  \citenamefont {Analytis}, \citenamefont {Chu}, \citenamefont {Liu},
  \citenamefont {Mo}, \citenamefont {Qi}, \citenamefont {Zhang}, \citenamefont
  {Lu}, \citenamefont {Dai}, \citenamefont {Fang}, \citenamefont {Zhang},
  \citenamefont {Fisher}, \citenamefont {Z.},\ and\ \citenamefont
  {Shen}}]{science178}%
  \BibitemOpen
  \bibfield  {author} {\bibinfo {author} {\bibfnamefont {Y.~L.}\ \bibnamefont
  {Chen}}, \bibinfo {author} {\bibfnamefont {J.~G.}\ \bibnamefont {Analytis}},
  \bibinfo {author} {\bibfnamefont {J.-H.}\ \bibnamefont {Chu}}, \bibinfo
  {author} {\bibfnamefont {Z.~K.}\ \bibnamefont {Liu}}, \bibinfo {author}
  {\bibfnamefont {S.-K.}\ \bibnamefont {Mo}}, \bibinfo {author} {\bibfnamefont
  {X.~L.}\ \bibnamefont {Qi}}, \bibinfo {author} {\bibfnamefont {H.~J.}\
  \bibnamefont {Zhang}}, \bibinfo {author} {\bibfnamefont {D.~H.}\ \bibnamefont
  {Lu}}, \bibinfo {author} {\bibfnamefont {X.}~\bibnamefont {Dai}}, \bibinfo
  {author} {\bibfnamefont {Z.}~\bibnamefont {Fang}}, \bibinfo {author}
  {\bibfnamefont {S.~C.}\ \bibnamefont {Zhang}}, \bibinfo {author}
  {\bibfnamefont {I.~R.}\ \bibnamefont {Fisher}}, \bibinfo {author}
  {\bibfnamefont {H.}~\bibnamefont {Z.}}, \ and\ \bibinfo {author}
  {\bibfnamefont {Z.-X.}\ \bibnamefont {Shen}},\ }\href {\doibase
  10.1126/science.1173034} {\bibfield  {journal} {\bibinfo  {journal}
  {Science}\ }\textbf {\bibinfo {volume} {325}},\ \bibinfo {pages} {178}
  (\bibinfo {year} {2009})}\BibitemShut {NoStop}%
\bibitem [{\citenamefont {Ren}\ \emph {et~al.}(2011)\citenamefont {Ren},
  \citenamefont {Taskin}, \citenamefont {Sasaki}, \citenamefont {Segawa},\ and\
  \citenamefont {Ando}}]{prb165311}%
  \BibitemOpen
  \bibfield  {author} {\bibinfo {author} {\bibfnamefont {Z.}~\bibnamefont
  {Ren}}, \bibinfo {author} {\bibfnamefont {A.~A.}\ \bibnamefont {Taskin}},
  \bibinfo {author} {\bibfnamefont {S.}~\bibnamefont {Sasaki}}, \bibinfo
  {author} {\bibfnamefont {K.}~\bibnamefont {Segawa}}, \ and\ \bibinfo {author}
  {\bibfnamefont {Y.}~\bibnamefont {Ando}},\ }\href {\doibase
  10.1103/PhysRevB.84.165311} {\bibfield  {journal} {\bibinfo  {journal} {Phys.
  Rev. B}\ }\textbf {\bibinfo {volume} {84}},\ \bibinfo {pages} {165311}
  (\bibinfo {year} {2011})}\BibitemShut {NoStop}%
\bibitem [{\citenamefont {Ren}\ \emph {et~al.}(2012)\citenamefont {Ren},
  \citenamefont {Taskin}, \citenamefont {Sasaki}, \citenamefont {Segawa},\ and\
  \citenamefont {Ando}}]{prb155301}%
  \BibitemOpen
  \bibfield  {author} {\bibinfo {author} {\bibfnamefont {Z.}~\bibnamefont
  {Ren}}, \bibinfo {author} {\bibfnamefont {A.~A.}\ \bibnamefont {Taskin}},
  \bibinfo {author} {\bibfnamefont {S.}~\bibnamefont {Sasaki}}, \bibinfo
  {author} {\bibfnamefont {K.}~\bibnamefont {Segawa}}, \ and\ \bibinfo {author}
  {\bibfnamefont {Y.}~\bibnamefont {Ando}},\ }\href {\doibase
  10.1103/PhysRevB.85.155301} {\bibfield  {journal} {\bibinfo  {journal} {Phys.
  Rev. B}\ }\textbf {\bibinfo {volume} {85}},\ \bibinfo {pages} {155301}
  (\bibinfo {year} {2012})}\BibitemShut {NoStop}%
\bibitem [{\citenamefont {Zhang}\ \emph {et~al.}(2011)\citenamefont {Zhang},
  \citenamefont {Chang}, \citenamefont {Zhang}, \citenamefont {Wen},
  \citenamefont {Feng}, \citenamefont {Li}, \citenamefont {Liu}, \citenamefont
  {He}, \citenamefont {Wang}, \citenamefont {Chen}, \citenamefont {Xue},
  \citenamefont {Ma},\ and\ \citenamefont {Wang}}]{nc574}%
  \BibitemOpen
  \bibfield  {author} {\bibinfo {author} {\bibfnamefont {J.}~\bibnamefont
  {Zhang}}, \bibinfo {author} {\bibfnamefont {C.-Z.}\ \bibnamefont {Chang}},
  \bibinfo {author} {\bibfnamefont {Z.}~\bibnamefont {Zhang}}, \bibinfo
  {author} {\bibfnamefont {J.}~\bibnamefont {Wen}}, \bibinfo {author}
  {\bibfnamefont {X.}~\bibnamefont {Feng}}, \bibinfo {author} {\bibfnamefont
  {K.}~\bibnamefont {Li}}, \bibinfo {author} {\bibfnamefont {M.}~\bibnamefont
  {Liu}}, \bibinfo {author} {\bibfnamefont {K.}~\bibnamefont {He}}, \bibinfo
  {author} {\bibfnamefont {L.}~\bibnamefont {Wang}}, \bibinfo {author}
  {\bibfnamefont {X.}~\bibnamefont {Chen}}, \bibinfo {author} {\bibfnamefont
  {Q.-K.}\ \bibnamefont {Xue}}, \bibinfo {author} {\bibfnamefont
  {X.}~\bibnamefont {Ma}}, \ and\ \bibinfo {author} {\bibfnamefont
  {Y.}~\bibnamefont {Wang}},\ }\href {\doibase 10.1038/ncomms1588} {\bibfield
  {journal} {\bibinfo  {journal} {Nat. Commun.}\ }\textbf {\bibinfo {volume}
  {2}},\ \bibinfo {pages} {574} (\bibinfo {year} {2011})}\BibitemShut {NoStop}%
\bibitem [{\citenamefont {Seidlmayer}(2010)}]{dis2010}%
  \BibitemOpen
  \bibfield  {author} {\bibinfo {author} {\bibfnamefont {S.}~\bibnamefont
  {Seidlmayer}},\ }\href@noop {} {\bibfield  {journal} {\bibinfo  {journal}
  {Dissertation Regensburg Universitaet}\ ,\ \bibinfo {pages} {1}} (\bibinfo
  {year} {2010})}\BibitemShut {NoStop}%
\bibitem [{\citenamefont {Hohenberg}\ and\ \citenamefont
  {Kohn}(1964)}]{prb864}%
  \BibitemOpen
  \bibfield  {author} {\bibinfo {author} {\bibfnamefont {P.}~\bibnamefont
  {Hohenberg}}\ and\ \bibinfo {author} {\bibfnamefont {W.}~\bibnamefont
  {Kohn}},\ }\href {\doibase 10.1103/PhysRev.136.B864} {\bibfield  {journal}
  {\bibinfo  {journal} {Phys. Rev.}\ }\textbf {\bibinfo {volume} {136}},\
  \bibinfo {pages} {B864} (\bibinfo {year} {1964})}\BibitemShut {NoStop}%
\bibitem [{\citenamefont {Perdew}\ \emph {et~al.}(1996)\citenamefont {Perdew},
  \citenamefont {Burke},\ and\ \citenamefont {Ernzerhof}}]{prl3865}%
  \BibitemOpen
  \bibfield  {author} {\bibinfo {author} {\bibfnamefont {J.~P.}\ \bibnamefont
  {Perdew}}, \bibinfo {author} {\bibfnamefont {K.}~\bibnamefont {Burke}}, \
  and\ \bibinfo {author} {\bibfnamefont {M.}~\bibnamefont {Ernzerhof}},\ }\href
  {\doibase 10.1103/PhysRevLett.77.3865} {\bibfield  {journal} {\bibinfo
  {journal} {Phys. Rev. Lett.}\ }\textbf {\bibinfo {volume} {77}},\ \bibinfo
  {pages} {3865} (\bibinfo {year} {1996})}\BibitemShut {NoStop}%
\bibitem [{\citenamefont {Kresse}\ and\ \citenamefont
  {Furthm\"{u}ller}(1996)}]{prb11169}%
  \BibitemOpen
  \bibfield  {author} {\bibinfo {author} {\bibfnamefont {G.}~\bibnamefont
  {Kresse}}\ and\ \bibinfo {author} {\bibnamefont {Furthm\"{u}ller}},\ }\href
  {\doibase 10.1103/PhysRevB.54.11169} {\bibfield  {journal} {\bibinfo
  {journal} {Phys. Rev. B}\ }\textbf {\bibinfo {volume} {54}},\ \bibinfo
  {pages} {11169} (\bibinfo {year} {1996})}\BibitemShut {NoStop}%
\bibitem [{\citenamefont {Togo}\ and\ \citenamefont {Tanaka}(2015)}]{sm1}%
  \BibitemOpen
  \bibfield  {author} {\bibinfo {author} {\bibfnamefont {A.}~\bibnamefont
  {Togo}}\ and\ \bibinfo {author} {\bibfnamefont {I.}~\bibnamefont {Tanaka}},\
  }\href {\doibase 10.1016/j.scriptamat.2015.07.021} {\bibfield  {journal}
  {\bibinfo  {journal} {Scr. Mater.}\ }\textbf {\bibinfo {volume} {108}},\
  \bibinfo {pages} {1} (\bibinfo {year} {2015})}\BibitemShut {NoStop}%
\bibitem [{\citenamefont {Arabbagheri}(2014)}]{dis2014}%
  \BibitemOpen
  \bibfield  {author} {\bibinfo {author} {\bibfnamefont {S.}~\bibnamefont
  {Arabbagheri}},\ }\href@noop {} {\bibfield  {journal} {\bibinfo  {journal}
  {Dissertation Regensburg Universitaet}\ ,\ \bibinfo {pages} {1}} (\bibinfo
  {year} {2014})}\BibitemShut {NoStop}%
\bibitem [{\citenamefont {Ji}\ \emph {et~al.}(2016)\citenamefont {Ji},
  \citenamefont {Pletikosi\'c}, \citenamefont {Gibson}, \citenamefont
  {Sahasrabudhe}, \citenamefont {Valla},\ and\ \citenamefont
  {Cava}}]{prb045315}%
  \BibitemOpen
  \bibfield  {author} {\bibinfo {author} {\bibfnamefont {H.}~\bibnamefont
  {Ji}}, \bibinfo {author} {\bibfnamefont {I.}~\bibnamefont {Pletikosi\'c}},
  \bibinfo {author} {\bibfnamefont {Q.~D.}\ \bibnamefont {Gibson}}, \bibinfo
  {author} {\bibfnamefont {G.}~\bibnamefont {Sahasrabudhe}}, \bibinfo {author}
  {\bibfnamefont {T.}~\bibnamefont {Valla}}, \ and\ \bibinfo {author}
  {\bibfnamefont {R.~J.}\ \bibnamefont {Cava}},\ }\href {\doibase
  10.1103/PhysRevB.93.045315} {\bibfield  {journal} {\bibinfo  {journal} {Phys.
  Rev. B}\ }\textbf {\bibinfo {volume} {93}},\ \bibinfo {pages} {045315}
  (\bibinfo {year} {2016})}\BibitemShut {NoStop}%
\bibitem [{\citenamefont {Teo}\ \emph {et~al.}(2008)\citenamefont {Teo},
  \citenamefont {Fu},\ and\ \citenamefont {Kane}}]{prb045426}%
  \BibitemOpen
  \bibfield  {author} {\bibinfo {author} {\bibfnamefont {J.~C.~Y.}\
  \bibnamefont {Teo}}, \bibinfo {author} {\bibfnamefont {L.}~\bibnamefont
  {Fu}}, \ and\ \bibinfo {author} {\bibfnamefont {C.~L.}\ \bibnamefont
  {Kane}},\ }\href {\doibase 10.1103/PhysRevB.78.045426} {\bibfield  {journal}
  {\bibinfo  {journal} {Phys. Rev. B}\ }\textbf {\bibinfo {volume} {78}},\
  \bibinfo {pages} {045426} (\bibinfo {year} {2008})}\BibitemShut {NoStop}%
\bibitem [{\citenamefont {Feng}\ \emph {et~al.}(2012)\citenamefont {Feng},
  \citenamefont {Wen}, \citenamefont {Zhou}, \citenamefont {Xiao},\ and\
  \citenamefont {Yao}}]{cpc1849}%
  \BibitemOpen
  \bibfield  {author} {\bibinfo {author} {\bibfnamefont {W.}~\bibnamefont
  {Feng}}, \bibinfo {author} {\bibfnamefont {J.}~\bibnamefont {Wen}}, \bibinfo
  {author} {\bibfnamefont {J.}~\bibnamefont {Zhou}}, \bibinfo {author}
  {\bibfnamefont {D.}~\bibnamefont {Xiao}}, \ and\ \bibinfo {author}
  {\bibfnamefont {Y.}~\bibnamefont {Yao}},\ }\href {\doibase
  10.1016/j.cpc.2012.04.001} {\bibfield  {journal} {\bibinfo  {journal}
  {Comput. Phys. Commun.}\ }\textbf {\bibinfo {volume} {183}},\ \bibinfo
  {pages} {1849} (\bibinfo {year} {2012})}\BibitemShut {NoStop}%
\bibitem [{\citenamefont {Hou}\ and\ \citenamefont {Wu}(2018)}]{arxiv07358}%
  \BibitemOpen
  \bibfield  {author} {\bibinfo {author} {\bibfnamefont {Y.}~\bibnamefont
  {Hou}}\ and\ \bibinfo {author} {\bibfnamefont {R.}~\bibnamefont {Wu}},\
  }\href {\doibase arXiv:1802.07358} {\bibfield  {journal} {\bibinfo  {journal}
  {arXiv preprint}\ ,\ \bibinfo {pages} {arXiv:1802.07358}} (\bibinfo {year}
  {2018})}\BibitemShut {NoStop}%
\bibitem [{\citenamefont {Zeisner}\ \emph {et~al.}(2019)\citenamefont
  {Zeisner}, \citenamefont {Alfonsov}, \citenamefont {Selter}, \citenamefont
  {Aswartham}, \citenamefont {Ghimire}, \citenamefont {Richter}, \citenamefont
  {van~den Brink}, \citenamefont {B$\ddot{u}$chner},\ and\ \citenamefont
  {Kataev}}]{arxiv02560}%
  \BibitemOpen
  \bibfield  {author} {\bibinfo {author} {\bibfnamefont {J.}~\bibnamefont
  {Zeisner}}, \bibinfo {author} {\bibfnamefont {A.}~\bibnamefont {Alfonsov}},
  \bibinfo {author} {\bibfnamefont {S.}~\bibnamefont {Selter}}, \bibinfo
  {author} {\bibfnamefont {S.}~\bibnamefont {Aswartham}}, \bibinfo {author}
  {\bibfnamefont {M.~P.}\ \bibnamefont {Ghimire}}, \bibinfo {author}
  {\bibfnamefont {M.}~\bibnamefont {Richter}}, \bibinfo {author} {\bibfnamefont
  {J.}~\bibnamefont {van~den Brink}}, \bibinfo {author} {\bibfnamefont
  {B.}~\bibnamefont {B$\ddot{u}$chner}}, \ and\ \bibinfo {author}
  {\bibfnamefont {V.}~\bibnamefont {Kataev}},\ }\href {\doibase
  arXiv:1810.02560} {\bibfield  {journal} {\bibinfo  {journal} {arXiv
  preprint}\ ,\ \bibinfo {pages} {arXiv:1810.02560}} (\bibinfo {year}
  {2019})}\BibitemShut {NoStop}%
\bibitem [{\citenamefont {Burch}\ \emph {et~al.}(2018)\citenamefont {Burch},
  \citenamefont {Mandrus},\ and\ \citenamefont {Park}}]{nature47}%
  \BibitemOpen
  \bibfield  {author} {\bibinfo {author} {\bibfnamefont {K.~S.}\ \bibnamefont
  {Burch}}, \bibinfo {author} {\bibfnamefont {D.}~\bibnamefont {Mandrus}}, \
  and\ \bibinfo {author} {\bibfnamefont {J.-G.}\ \bibnamefont {Park}},\ }\href
  {\doibase 10.1038/s41586-018-0631-z} {\bibfield  {journal} {\bibinfo
  {journal} {Nature}\ }\textbf {\bibinfo {volume} {563}},\ \bibinfo {pages}
  {47} (\bibinfo {year} {2018})}\BibitemShut {NoStop}%
\bibitem [{\citenamefont {Song}\ \emph {et~al.}(2018)\citenamefont {Song},
  \citenamefont {Cai}, \citenamefont {Tu}, \citenamefont {Zhang}, \citenamefont
  {Huang}, \citenamefont {Wilson}, \citenamefont {Seyler}, \citenamefont {Zhu},
  \citenamefont {Taniguchi}, \citenamefont {Watanabe}, \citenamefont {McGuire},
  \citenamefont {Cobden}, \citenamefont {Xiao}, \citenamefont {Yao},\ and\
  \citenamefont {Xu}}]{science1214}%
  \BibitemOpen
  \bibfield  {author} {\bibinfo {author} {\bibfnamefont {T.}~\bibnamefont
  {Song}}, \bibinfo {author} {\bibfnamefont {X.}~\bibnamefont {Cai}}, \bibinfo
  {author} {\bibfnamefont {M.~W.-Y.}\ \bibnamefont {Tu}}, \bibinfo {author}
  {\bibfnamefont {X.}~\bibnamefont {Zhang}}, \bibinfo {author} {\bibfnamefont
  {B.}~\bibnamefont {Huang}}, \bibinfo {author} {\bibfnamefont
  {N.}~\bibnamefont {Wilson}}, \bibinfo {author} {\bibfnamefont {K.~L.}\
  \bibnamefont {Seyler}}, \bibinfo {author} {\bibfnamefont {L.}~\bibnamefont
  {Zhu}}, \bibinfo {author} {\bibfnamefont {T.}~\bibnamefont {Taniguchi}},
  \bibinfo {author} {\bibfnamefont {K.}~\bibnamefont {Watanabe}}, \bibinfo
  {author} {\bibfnamefont {M.~A.}\ \bibnamefont {McGuire}}, \bibinfo {author}
  {\bibfnamefont {D.~H.}\ \bibnamefont {Cobden}}, \bibinfo {author}
  {\bibfnamefont {D.}~\bibnamefont {Xiao}}, \bibinfo {author} {\bibfnamefont
  {W.}~\bibnamefont {Yao}}, \ and\ \bibinfo {author} {\bibfnamefont
  {X.}~\bibnamefont {Xu}},\ }\href {\doibase 10.1126/science.aar4851}
  {\bibfield  {journal} {\bibinfo  {journal} {Science}\ }\textbf {\bibinfo
  {volume} {360}},\ \bibinfo {pages} {1214} (\bibinfo {year}
  {2018})}\BibitemShut {NoStop}%
\bibitem [{\citenamefont {Klein}\ \emph {et~al.}(2018)\citenamefont {Klein},
  \citenamefont {MacNeill}, \citenamefont {Lado}, \citenamefont {Soriano},
  \citenamefont {Navarro-Moratalla}, \citenamefont {Watanabe}, \citenamefont
  {Taniguchi}, \citenamefont {Manni}, \citenamefont {P.}, \citenamefont
  {Fern\'{a}ndez-Rossier},\ and\ \citenamefont
  {Jarillo-Herrero}}]{science1218}%
  \BibitemOpen
  \bibfield  {author} {\bibinfo {author} {\bibfnamefont {D.~R.}\ \bibnamefont
  {Klein}}, \bibinfo {author} {\bibfnamefont {D.}~\bibnamefont {MacNeill}},
  \bibinfo {author} {\bibfnamefont {J.~L.}\ \bibnamefont {Lado}}, \bibinfo
  {author} {\bibfnamefont {D.}~\bibnamefont {Soriano}}, \bibinfo {author}
  {\bibfnamefont {E.}~\bibnamefont {Navarro-Moratalla}}, \bibinfo {author}
  {\bibfnamefont {K.}~\bibnamefont {Watanabe}}, \bibinfo {author}
  {\bibfnamefont {T.}~\bibnamefont {Taniguchi}}, \bibinfo {author}
  {\bibfnamefont {S.}~\bibnamefont {Manni}}, \bibinfo {author} {\bibfnamefont
  {C.}~\bibnamefont {P.}}, \bibinfo {author} {\bibfnamefont {J.}~\bibnamefont
  {Fern\'{a}ndez-Rossier}}, \ and\ \bibinfo {author} {\bibfnamefont
  {P.}~\bibnamefont {Jarillo-Herrero}},\ }\href {\doibase
  10.1126/science.aar3617} {\bibfield  {journal} {\bibinfo  {journal}
  {Science}\ }\textbf {\bibinfo {volume} {360}},\ \bibinfo {pages} {1218}
  (\bibinfo {year} {2018})}\BibitemShut {NoStop}%
\bibitem [{\citenamefont {Huang}\ \emph {et~al.}(2017)\citenamefont {Huang},
  \citenamefont {Clark}, \citenamefont {Navarro-Moratalla}, \citenamefont
  {Klein}, \citenamefont {Cheng}, \citenamefont {Seyler}, \citenamefont
  {Zhong}, \citenamefont {Schmidgall}, \citenamefont {McGuire}, \citenamefont
  {Cobden}, \citenamefont {Yao}, \citenamefont {Xiao}, \citenamefont
  {Jarillo-Herrero},\ and\ \citenamefont {Xu}}]{nature270}%
  \BibitemOpen
  \bibfield  {author} {\bibinfo {author} {\bibfnamefont {B.}~\bibnamefont
  {Huang}}, \bibinfo {author} {\bibfnamefont {G.}~\bibnamefont {Clark}},
  \bibinfo {author} {\bibfnamefont {E.}~\bibnamefont {Navarro-Moratalla}},
  \bibinfo {author} {\bibfnamefont {D.~R.}\ \bibnamefont {Klein}}, \bibinfo
  {author} {\bibfnamefont {R.}~\bibnamefont {Cheng}}, \bibinfo {author}
  {\bibfnamefont {K.~L.}\ \bibnamefont {Seyler}}, \bibinfo {author}
  {\bibfnamefont {D.}~\bibnamefont {Zhong}}, \bibinfo {author} {\bibfnamefont
  {E.}~\bibnamefont {Schmidgall}}, \bibinfo {author} {\bibfnamefont {M.~A.}\
  \bibnamefont {McGuire}}, \bibinfo {author} {\bibfnamefont {D.~H.}\
  \bibnamefont {Cobden}}, \bibinfo {author} {\bibfnamefont {W.}~\bibnamefont
  {Yao}}, \bibinfo {author} {\bibfnamefont {D.}~\bibnamefont {Xiao}}, \bibinfo
  {author} {\bibfnamefont {P.}~\bibnamefont {Jarillo-Herrero}}, \ and\ \bibinfo
  {author} {\bibfnamefont {X.}~\bibnamefont {Xu}},\ }\href {\doibase
  10.1038/nature22391} {\bibfield  {journal} {\bibinfo  {journal} {Nature}\
  }\textbf {\bibinfo {volume} {546}},\ \bibinfo {pages} {270} (\bibinfo {year}
  {2017})}\BibitemShut {NoStop}%
\bibitem [{\citenamefont {Ji}\ \emph {et~al.}(2013)\citenamefont {Ji},
  \citenamefont {Stokes}, \citenamefont {Alegria}, \citenamefont {Blomberg},
  \citenamefont {Tanatar}, \citenamefont {Reijnders}, \citenamefont {Schoop},
  \citenamefont {Liang}, \citenamefont {Prozorov}, \citenamefont {Burch},
  \citenamefont {Ong}, \citenamefont {Petta},\ and\ \citenamefont
  {Cava}}]{jap114907}%
  \BibitemOpen
  \bibfield  {author} {\bibinfo {author} {\bibfnamefont {H.}~\bibnamefont
  {Ji}}, \bibinfo {author} {\bibfnamefont {R.~A.}\ \bibnamefont {Stokes}},
  \bibinfo {author} {\bibfnamefont {L.~D.}\ \bibnamefont {Alegria}}, \bibinfo
  {author} {\bibfnamefont {E.~C.}\ \bibnamefont {Blomberg}}, \bibinfo {author}
  {\bibfnamefont {M.~A.}\ \bibnamefont {Tanatar}}, \bibinfo {author}
  {\bibfnamefont {A.}~\bibnamefont {Reijnders}}, \bibinfo {author}
  {\bibfnamefont {L.~M.}\ \bibnamefont {Schoop}}, \bibinfo {author}
  {\bibfnamefont {T.}~\bibnamefont {Liang}}, \bibinfo {author} {\bibfnamefont
  {R.}~\bibnamefont {Prozorov}}, \bibinfo {author} {\bibfnamefont {K.~S.}\
  \bibnamefont {Burch}}, \bibinfo {author} {\bibfnamefont {N.~P.}\ \bibnamefont
  {Ong}}, \bibinfo {author} {\bibfnamefont {J.~R.}\ \bibnamefont {Petta}}, \
  and\ \bibinfo {author} {\bibfnamefont {R.~J.}\ \bibnamefont {Cava}},\ }\href
  {\doibase 10.1063/1.4822092} {\bibfield  {journal} {\bibinfo  {journal} {J.
  Appl. Phys.}\ }\textbf {\bibinfo {volume} {114}},\ \bibinfo {pages} {114907}
  (\bibinfo {year} {2013})}\BibitemShut {NoStop}%
\bibitem [{\citenamefont {Chong}\ \emph {et~al.}(2018)\citenamefont {Chong},
  \citenamefont {Han}, \citenamefont {Nagaoka}, \citenamefont {Tsuchikawa},
  \citenamefont {Liu}, \citenamefont {Liu}, \citenamefont {Vardeny},
  \citenamefont {Pesin}, \citenamefont {Lee}, \citenamefont {Sparks},\ and\
  \citenamefont {Deshpande}}]{nl8047}%
  \BibitemOpen
  \bibfield  {author} {\bibinfo {author} {\bibfnamefont {S.~K.}\ \bibnamefont
  {Chong}}, \bibinfo {author} {\bibfnamefont {K.~B.}\ \bibnamefont {Han}},
  \bibinfo {author} {\bibfnamefont {A.}~\bibnamefont {Nagaoka}}, \bibinfo
  {author} {\bibfnamefont {R.}~\bibnamefont {Tsuchikawa}}, \bibinfo {author}
  {\bibfnamefont {R.}~\bibnamefont {Liu}}, \bibinfo {author} {\bibfnamefont
  {H.}~\bibnamefont {Liu}}, \bibinfo {author} {\bibfnamefont {Z.~V.}\
  \bibnamefont {Vardeny}}, \bibinfo {author} {\bibfnamefont {D.~A.}\
  \bibnamefont {Pesin}}, \bibinfo {author} {\bibfnamefont {C.}~\bibnamefont
  {Lee}}, \bibinfo {author} {\bibfnamefont {T.~D.}\ \bibnamefont {Sparks}}, \
  and\ \bibinfo {author} {\bibfnamefont {V.~V.}\ \bibnamefont {Deshpande}},\
  }\href {\doibase 10.1021/acs.nanolett.8b04291} {\bibfield  {journal}
  {\bibinfo  {journal} {Nano Lett.}\ }\textbf {\bibinfo {volume} {18}},\
  \bibinfo {pages} {8047} (\bibinfo {year} {2018})}\BibitemShut {NoStop}%
\bibitem [{\citenamefont {Zhang}\ \emph {et~al.}(2010)\citenamefont {Zhang},
  \citenamefont {Yu}, \citenamefont {Zhang}, \citenamefont {Dai},\ and\
  \citenamefont {Fang}}]{njp065013}%
  \BibitemOpen
  \bibfield  {author} {\bibinfo {author} {\bibfnamefont {W.}~\bibnamefont
  {Zhang}}, \bibinfo {author} {\bibfnamefont {R.}~\bibnamefont {Yu}}, \bibinfo
  {author} {\bibfnamefont {H.-J.}\ \bibnamefont {Zhang}}, \bibinfo {author}
  {\bibfnamefont {X.}~\bibnamefont {Dai}}, \ and\ \bibinfo {author}
  {\bibfnamefont {Z.}~\bibnamefont {Fang}},\ }\href@noop {} {\bibfield
  {journal} {\bibinfo  {journal} {New J. Phys.}\ }\textbf {\bibinfo {volume}
  {12}},\ \bibinfo {pages} {065103} (\bibinfo {year} {2010})}\BibitemShut
  {NoStop}%
\end{thebibliography}%

\end{document}


\beginsupplement
\section{Supplementary Materials}

\begin{figure*}[htp!]
\includegraphics [width=12cm]{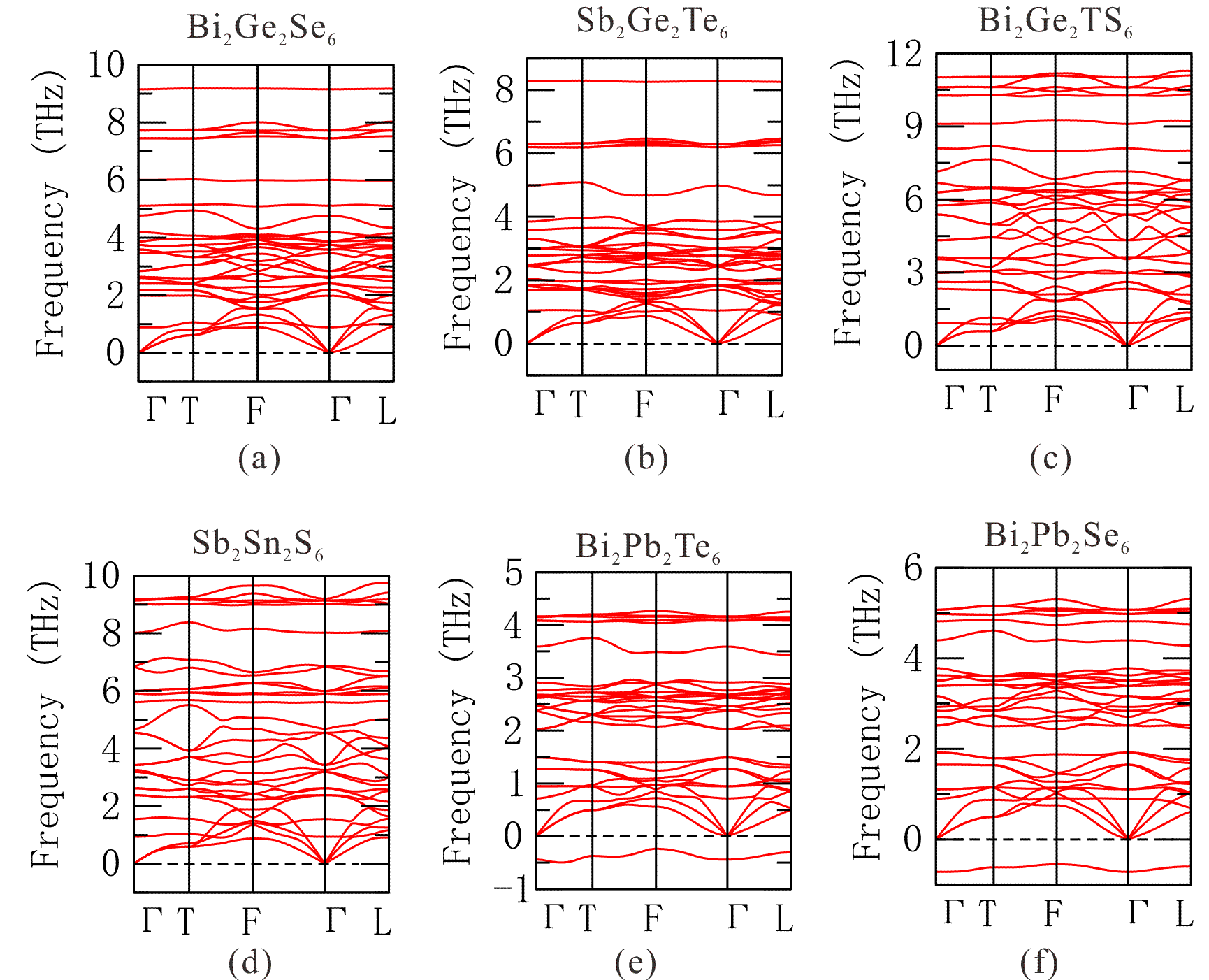}
\caption{ Phonon dispersions of more 3D V$_2$IV$_2$VI$_6$ crystals. }
\label{spho}
\end{figure*}

\clearpage

\begin{figure*}
\includegraphics [width=12cm]{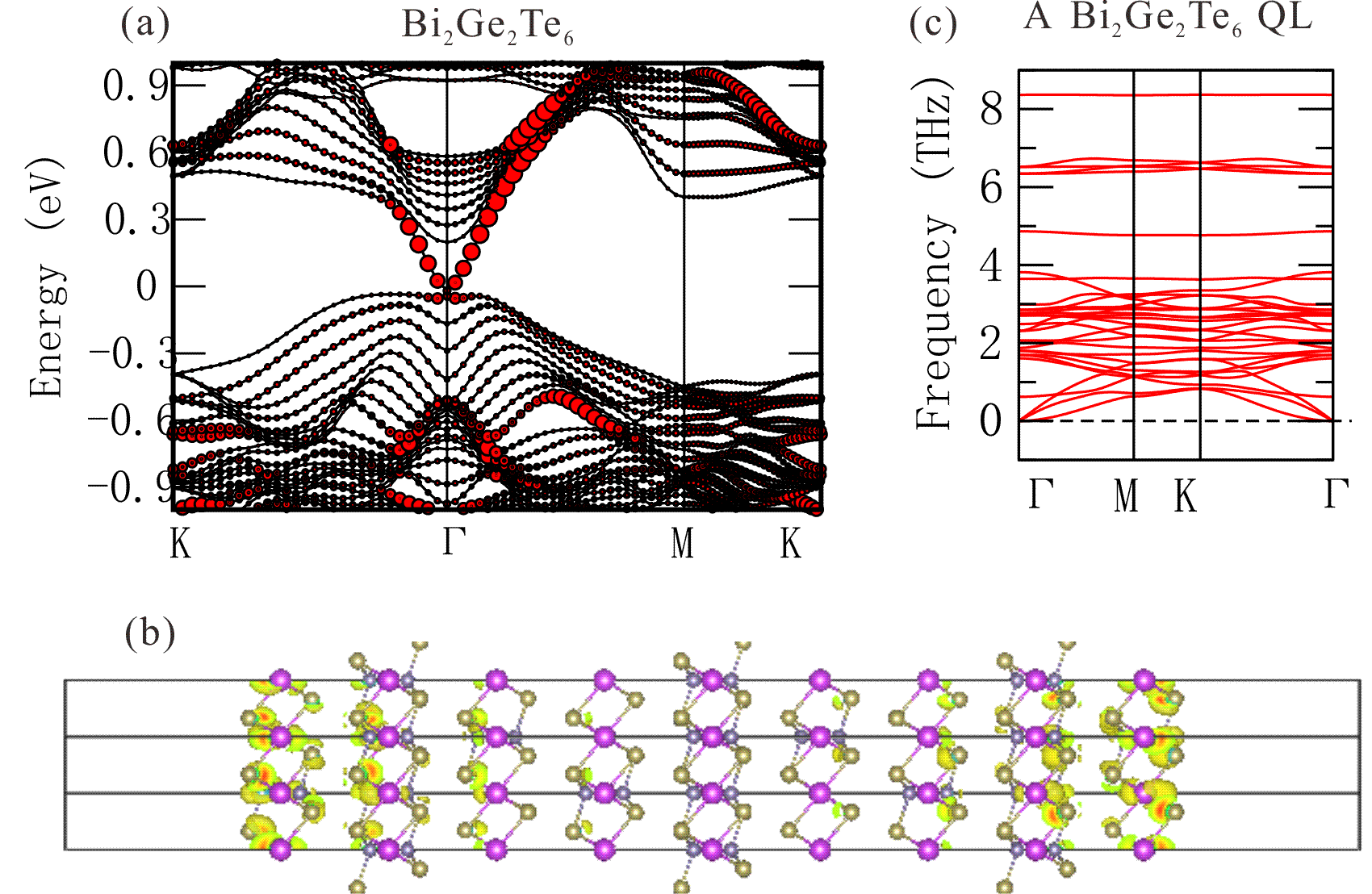}
\caption{ (a) Band structure of a Bi$_2$Ge$_2$Te$_6$ slab. Red symbols denote contributions of surface atoms. (b) Projected wavefunctions of the Dirac point in (a). (c) Phonon dispersions of a Bi$_2$Ge$_2$Te$_6$ QL. }
\label{sslab}
\end{figure*}